\documentclass{aa}  
\usepackage[export]{adjustbox}
\usepackage{scalerel}
\usepackage{lscape}
\newcommand\HI{H\protect\scaleto{$I$}{1.2ex}}
\newcommand\hi{H\protect\scaleto{$I$}{1.2ex}}

\newcommand\htwo{H{$_2$}\protect\scaleto{1.2ex}}
\newcommand{\kms}{$\,$km$\,$s$^{-1}$}

\newcommand\mJb{mJy beam$^{-1}$}
\newcommand\vbary{$v_{\text{H\protect\scaleto{$I$}{1.2ex}}}$}
\newcommand\vspe{$v_\mathrm{opt}$}
\newcommand\mhi{$M_{\text{H\protect\scaleto{$I$}{1.2ex}}}$}
\newcommand\mhtwo{$M_{\text{H}_2}$}

\newcommand\nhi{$N_{\text{H\protect\scaleto{$I$}{1.2ex}}}$}
\newcommand\mst{$M_\star$}
\usepackage{caption}
\usepackage{subcaption} 
\usepackage{xcolor}
\usepackage{booktabs}
\usepackage{graphbox,graphicx}
\usepackage{txfonts}
\usepackage{hyperref}

\usepackage[toc,page]{appendix}

\begin{document} 

   \title{A blind ATCA \hi{} survey of the Fornax galaxy cluster}
   \subtitle{Properties of the \hi{} detections}
   \author{A.~Loni
          \inst{1,2}
          \and
          P.~Serra\inst{1}
          \and
          D.~Kleiner\inst{1}
          \and
          L.~Cortese\inst{3,4}
          \and
          B.~Catinella\inst{3,4}
          \and
          B.~Koribalski\inst{5}
          \and
          T.~H.~Jarrett\inst{6}    
          \and
          D.~Cs.~Molnar\inst{1}
          \and
          T.~A.~Davis\inst{7}
         \and 
          E.~Iodice\inst{8}
          \and          
          K.~Lee-Waddell\inst{5}
          \and         
          F.~Loi\inst{1} 
          \and
          F.~M.~Maccagni\inst{1}          
          \and
          R.~Peletier\inst{9}
          \and
          A.~Popping\inst{3,10}
          \and
          M.~Ramatsoku\inst{1,11}
          \and
          M.~W~.L.~Smith\inst{7}
          \and
          N.~Zabel\inst{9}
          }

   \institute{INAF - Osservatorio Astronomico di Cagliari, Via della Scienza 5, 09047, Selargius, CA, Italy \\    \email{alessandro.loni@inaf.it}
         \and 
        Dipartimento di Fisica, Università di Cagliari, Cittadella Universitaria, 09042, Monserrato, Italy
         \and 
         International Centre for Radio Astronomy Research (ICRAR), The University of Western Australia, 35 Stirling Hwy, Crawley, WA 6009, Australia
        \and 
         ARC Centre of Excellence for All-Sky Astrophysics in 3 Dimensions (ASTRO3D)
        \and 
        CSIRO Astronomy and Space Science, Australia Telescope National Facility PO Box 76, Epping, NSW 1710, Australia
        \and 
        Department of Astronomy, University of Cape Town, Private Bag X3, Rondebosch 7701, South Africa         
         \and
         School of Physics and Astronomy, Cardiff University, Queens Buildings The Parade, Cardiff CF24 3AA, UK 
         \and     
         INAF-Astronomical observatory of Capodimonte, via Moiariello 16, Naples 80131, Italy
        \and 
        Kapteyn Astronomical Institute, University of Groningen, PO Box 800, 9700 AV Groningen, The Netherlands
         \and 
         Australian Research Council, Centre of Excellence for All-sky Astrophysics (CAASTRO), Australia
         \and 
         Department of Physics and Electronics, Rhodes University, PO Box 94, Makhanda, 6140, South Africa
         }

   \date{Received October 30, 2020; accepted January 26, 2021}

  \abstract
  {We present the first interferometric blind \hi{} survey of the Fornax galaxy cluster, which covers an area of 15~deg$^2$ out to the cluster virial radius. The survey has a spatial and velocity resolution of 67\arcsec~$\times~$95\arcsec ($\sim6\times9$~kpc at the Fornax cluster distance of 20~Mpc) and 6.6~\kms{} and a 3$\sigma$ sensitivity of \nhi{}~$\sim$2$~\times$~10$^{19}$~cm$^{-2}$ and \mhi{}~$\sim$2$~\times$~10$^7$~M$_\odot$, respectively. 
  We detect 16 galaxies out of roughly 200 spectroscopically confirmed Fornax cluster members. The detections cover about three orders of magnitude in \hi{} mass, from $8\times 10^6$ to $1.5\times 10^{10}$~M$_\odot$.  They avoid the central, virialised region of the cluster both on the sky and in projected phase-space, showing that they are recent arrivals and that, in Fornax, \hi{} is lost within a crossing time, $\sim2$~Gyr. Half of these galaxies exhibit a disturbed \hi{} morphology, including several cases of asymmetries, tails, offsets between \hi{} and optical centres, and a case of a truncated \hi{} disc.
  This suggests that these recent arrivals have been interacting with other galaxies, the large-scale potential or the intergalactic medium, within or on their way to Fornax. As a whole, our Fornax \hi{} detections are \hi{}-poorer and form stars at a lower rate than non-cluster galaxies in the same \mst{} range. This is particularly evident at \mst{}$\lesssim10^9$~M$_\odot$, indicating that low mass galaxies are more strongly affected throughout their infall towards the cluster. The \mhi{}/\mst{} ratio of Fornax galaxies is comparable to that in the Virgo cluster. At fixed \mst{}, our \hi{} detections follow the non-cluster relation between \mhi{} and the star formation rate, and we argue that this implies that thus far they have lost their \hi{} on a timescale $\gtrsim 1\mathrm{-}2$~Gyr. Deeper inside the cluster \hi{} removal is likely to proceed faster, as confirmed by a population of \hi{}-undetected but \htwo{0pt}-detected star-forming galaxies. Overall, based on ALMA data, we find a large scatter in \htwo{0pt}-to-\hi{} mass ratio, with several galaxies showing an unusually high ratio that is probably caused by faster \hi{} removal.
  Finally, we identify an \hi{}-rich subgroup of possible interacting galaxies dominated by NGC~1365, where pre-processing is likey to have taken place.} 


   \maketitle

\section{Introduction}
\label{sec:intro}
    It is known that the evolution of galaxies is faster in denser environments \citep{2001Diaferio} and that, as a consequence, the relative abundance of red early-type galaxies and blue late-type galaxies changes with environment density
    \citep[]{1931ApJ....74...43H, 1974ApJ...194....1O, dressler1980galaxy}. 
    In this context, galaxy clusters are the most extreme environments within which galaxies evolve.
    They are characterised by a high number density of galaxies and by the presence of a dense intra-cluster medium (ICM), which galaxies move through. Thus, in clusters, both hydrodynamical and gravitational interactions such as ram-pressure stripping and tidal interactions, respectively, are likely to happen (\citealt{1972ApJ...176....1G, 1972ApJ...178..623T}), as well as mergers between galaxies (at least in the cluster outskirts - \citealt{2012ApJS..202....8S, 2018ApJS..237...14O}). These interactions deplete the cold gas reservoirs of galaxies and, therefore, affect their star formation activity. 
    The balance between these types of interactions depends both on galaxy properties -- including their orbits -- and on cluster properties, such as the number density of galaxies and the ICM density. Therefore, we expect galaxy evolution to proceed differently in different clusters.

    The aim of this work is to study galaxy evolution in the Fornax cluster, which is
    a low mass cluster (7$\times10^{13}$~M$_\odot$ within a radius of 1.4~Mpc - twice the virial radius, $R_\mathrm{vir}$; \citealt{drinkwater2001substructure}), located in the southern sky at a distance of 20~Mpc \citep{blakeslee2009acs}. Fornax is thus the second closest galaxy cluster to us after Virgo. 
    Fornax has roughly 200 spectroscopically confirmed galaxies within $R_\mathrm{vir}\sim$700~kpc \citep{Maddox2019}. NGC~1399 is the central dominant galaxy, which coincides with the peak of the X-ray emission \citep{paolillo2002xray}, and  the majority of the massive galaxies in the cluster central region are of an early morphological type. Thus, Fornax shows a more dynamically evolved state than Virgo \citep{Grillmair1994_ngc1399,JAndres2007_ACSfornax}, but its growth is not over at all. 

    The Fornax region includes two main substructures \citep{drinkwater2001substructure}.  The first is the cluster itself centred on NGC~1399, whose interaction with NGC~1404 is revealed by a perturbed ICM distribution (e.g. \citealt{Sheardown_Fornaxhistory}). Here, three well defined groups of galaxies with different light and colour distributions, kinematics, and stellar populations were found by combining the Fornax Deep Survey imaging \citep{2019A&A...623A...1I} with Fornax 3D spectroscopy \citep{iodiceFDS2019} in the two-dimensional projected phase space: the core, the north-south clump and the infalling galaxies (see Fig.~7 in \citealt{iodiceFDS2019}).
    The core is still dominated by NGC~1399, which is one of only two slow-rotators inside the virial radius (NGC~1427 is the other one, located on the east side of the cluster).
    The NS-clump is located in the high-density region of the cluster (within $0.4 R_\mathrm{vir} \sim 0.3$~Mpc in projection), where the X-ray emission is still bright. It hosts the reddest and most metal-rich galaxies, all of them fast-rotating early-type galaxies. The bulk of the gravitational interactions between Fornax galaxies as well as most of the intra-cluster baryons (i.e. diffuse light, globular clusters and planetary nebulae) \citep{2018A&A...611A..93C,2018MNRAS.477.1880S,2019A&A...623A...1I} are found in this NS clump, where galaxy growth is still ongoing through the accretion of mass onto galaxies' outer regions \citep{2020A&A...639A..14S}. The third group of objects in the cluster includes the infalling galaxies, which are distributed nearly symmetrically around the core in the low-density region outside $\sim0.4 R_\mathrm{vir} \sim 0.3$~Mpc in projection. The majority of these galaxies are late-types with ongoing star formation. Most of them exhibit signs of interaction with the environment and/or minor mergers in the form of tidal tails and disturbed molecular gas \citep{zabel,FDS_RAJ}. Previous works had also shown that numerous dwarf galaxies are currently falling into the centre of the cluster  \citep{Drinkwater2001_evolutionofFornaxGal,Schroder2001,Waugh2002}. 
    Finally, the other major structure in the Fornax volume is the infalling group centred on NGC~1316 (Fornax~A), located $\sim$ 1.5 Mpc south-west 
    of the cluster  centre and experiencing ongoing interactions between its members \citep{Schweizer_opticalFornaxA,Horellou2001_fornaxA,FDSIodice_2017_fornaxA,Paolo2019_fornaxA}.
    
    The observation of galaxies' atomic hydrogen through the 21 cm wavelength emission line (\hi{}) gives us information on the evolutionary state of galaxies as well as a global picture of the cluster.
    The evolution of galaxies depends on the evolution of their atomic hydrogen reservoir, which is the primary reservoir of fuel for star formation. 
    Since it typically extends to the outskirts of galaxies, \hi{} gas is the first component that is 
    affected by tidal interactions, ram-pressure stripping and mergers. Indeed, cluster galaxies are usually deficient in atomic hydrogen with respect to non-cluster galaxies \citep{giovanelli1983AJ.....88..881G, Haynes1986ApJ...306..466H, boselligavazzi}.
    \hi{} is, therefore, a crucial observable for understanding galaxy evolution in dense environments \citep{HughesCortese2009_environm, ChungVIVA2009}.
    
    \hi{} emission in the Fornax cluster was studied in several works. 
    \cite{Bureau1996} used the Parkes radio telescope to measure the amount of atomic hydrogen in 21 undisturbed galaxies with morphologies S0$/$a or later, located within 6~deg from NGC~1399 and with c\textit{z}~$\leq$~2520~\kms{}.  
    The eight galaxies within $R_\mathrm{vir}$ did not show any peculiar value in the \mhi{}/\textit{I}$-$band infrared luminosity ratio with respect to the rest of the sample.
    The same ratio evaluated for Ursa Major galaxies, a lower density environment than Fornax, led them to conclude that the observed Fornax galaxies are not \hi{} deficient.
    
    The first blind \hi{} survey of Fornax was carried out by \cite{Barnes1997fornaxcluster}, also with the Parkes telescope.  
    They detected \hi{} in eight galaxies within an area of  8~$\times$~8~deg$^2$. Of those, two galaxies are within the cluster $R_\mathrm{vir}$: ESO~358G-063 and NGC~1365. 
    Thanks to the survey sensitivity, they excluded the existence of a significant population of optically undetected \hi{} clouds with \hi{} mass greater than 10$^8$ M$_{\odot}$.

    The number of \hi{} detections within the Fornax central region increased with the targeted survey by \cite{Schroder2001}.
    They found \hi{} in 37 out of 66 galaxies. Of those, 14 are within $R_\mathrm{vir}$, while the rest lies within 5 deg from the centre of the cluster. 
    They found a lack of \hi{} in the cluster centre, and measured the \hi{} deficiency parameter to be $0.38\pm0.09$ (as defined in \citealt{SolanesHImassfunc}). This shows a modest \hi{} depletion in Fornax (usually, galaxies with \hi{} deficiency parameter > 0.3 are considered to be \hi{} deficient, e.g. \citealt{Dressler1986_vdisp,Solanes2001_DEF}).  
    Furthermore, the \mhi{}$-$to$-$blue light ratio of Fornax galaxies - mean value ($0.68\pm 0.15$)~M$_{\odot}/$L$_{\odot}$  - is a factor 1.7 lower than in the field. 
    They also showed that the velocity dispersion of the sample of \hi{} deficient galaxies is lower than that of the remaining \hi{} detected galaxies. This difference in velocity dispersion agrees with the deficient galaxies having more radial orbits (as shown in \citealt{Dressler1986_vdisp}), which makes them good candidates for ram-pressure stripping (as discussed in \citealt{Solanes2001_DEF}). 

   \begin{figure*}
        \centering
        \includegraphics[width=180mm]{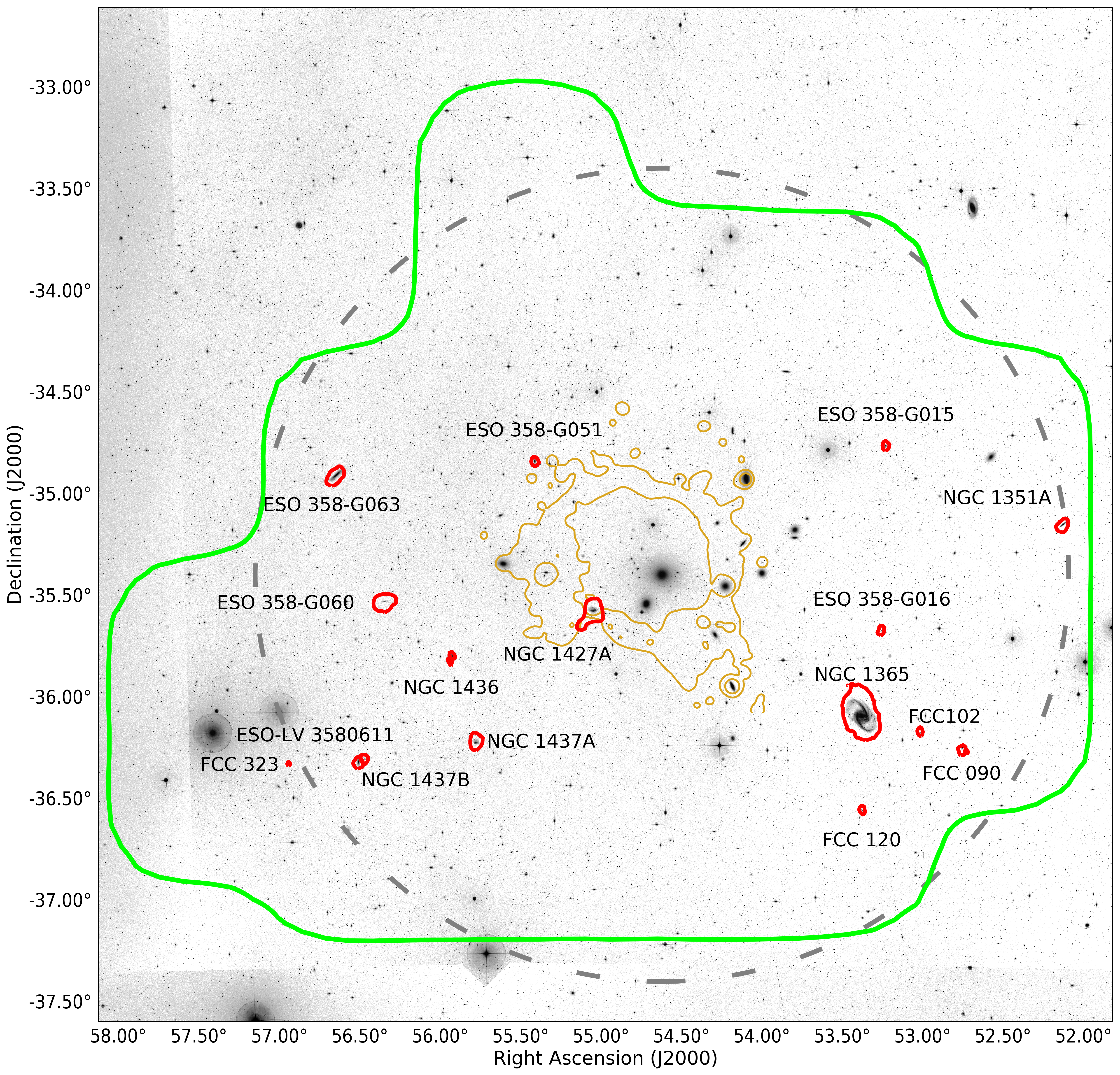} 
        \caption{Layout and detections of our ATCA \hi{} survey. The green outline includes the 15-deg$^2$ region where the average noise is 2.8~\mJb{} (see section \ref{sec:obs}). The red contours represent the lowest reliable \hi{} column density -- 3$\sigma$ over 25~\kms{} -- of our 16 detections. The grey dashed circle is $R_\mathrm{vir}$. The background optical image comes from the Digital Sky Survey (blue band). The yellow contours show the X-ray emission in the Fornax cluster detected with XMM-Newton \citep{2013ApJ...764...46F} and convolved with a 3~arcmin FWHM gaussian kernel. These contours are spaced by a factor of 2, with the lowest level at 3.7 counts deg$^{-2}$ s$^{- 1}$.}
        \label{fig:foot}
    \end{figure*}{}

    \begin{figure*}[t]
    \centering
    \includegraphics[width=170mm]{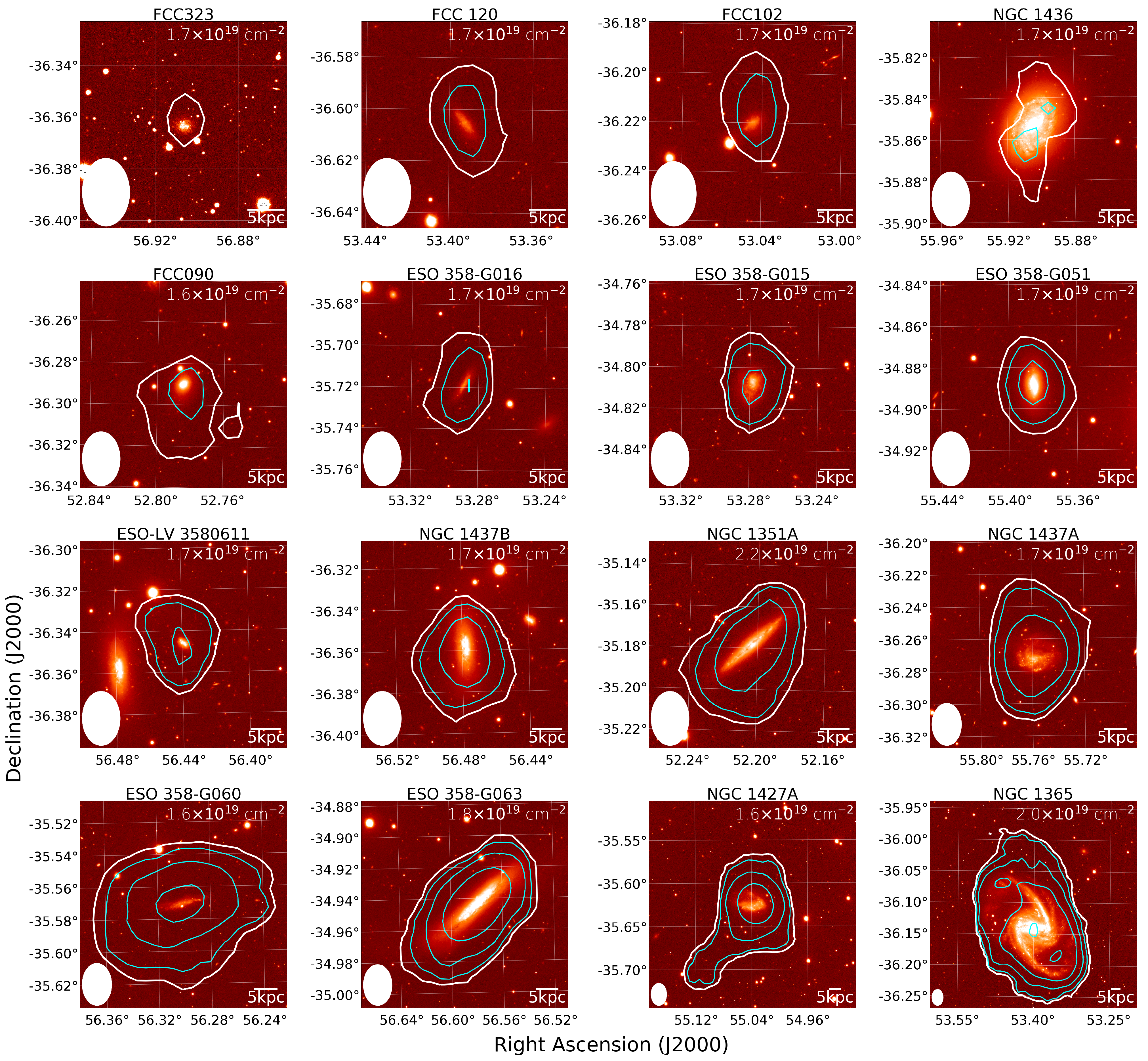}
  \caption{ATCA \hi{} contours overlaid on an optical image for all our \hi{} detections, sorted according to increasing \hi{} mass. The \textit{g}-band optical images come from the Fornax Deep Survey (\citealp{FDSIodice2016}; \citealp{venhola2018fds}; \citealp{Reynier_2020arXiv200812633P}) for all galaxies except FCC~323, whose \textit{g}-band optical image comes from the DESI Legacy Imaging Surveys, DR8 release, \citep{DR8_2019AJ....157..168D}. 
  In each panel we show the 3$\sigma$ column density sensitivity - values reported in the top right corner - with white colour, while cyan contours represent steps of 3$^n$ from it (n = 0, 1, 2, ...). We show the PSF on the bottom-left corner of each panel, and a 5~kpc scale bar in the bottom-right corner.}
  \label{fig:allmorph}
  \end{figure*}{} 
  
      \begin{figure*}[t]
        \centering
        \includegraphics[width=170mm]{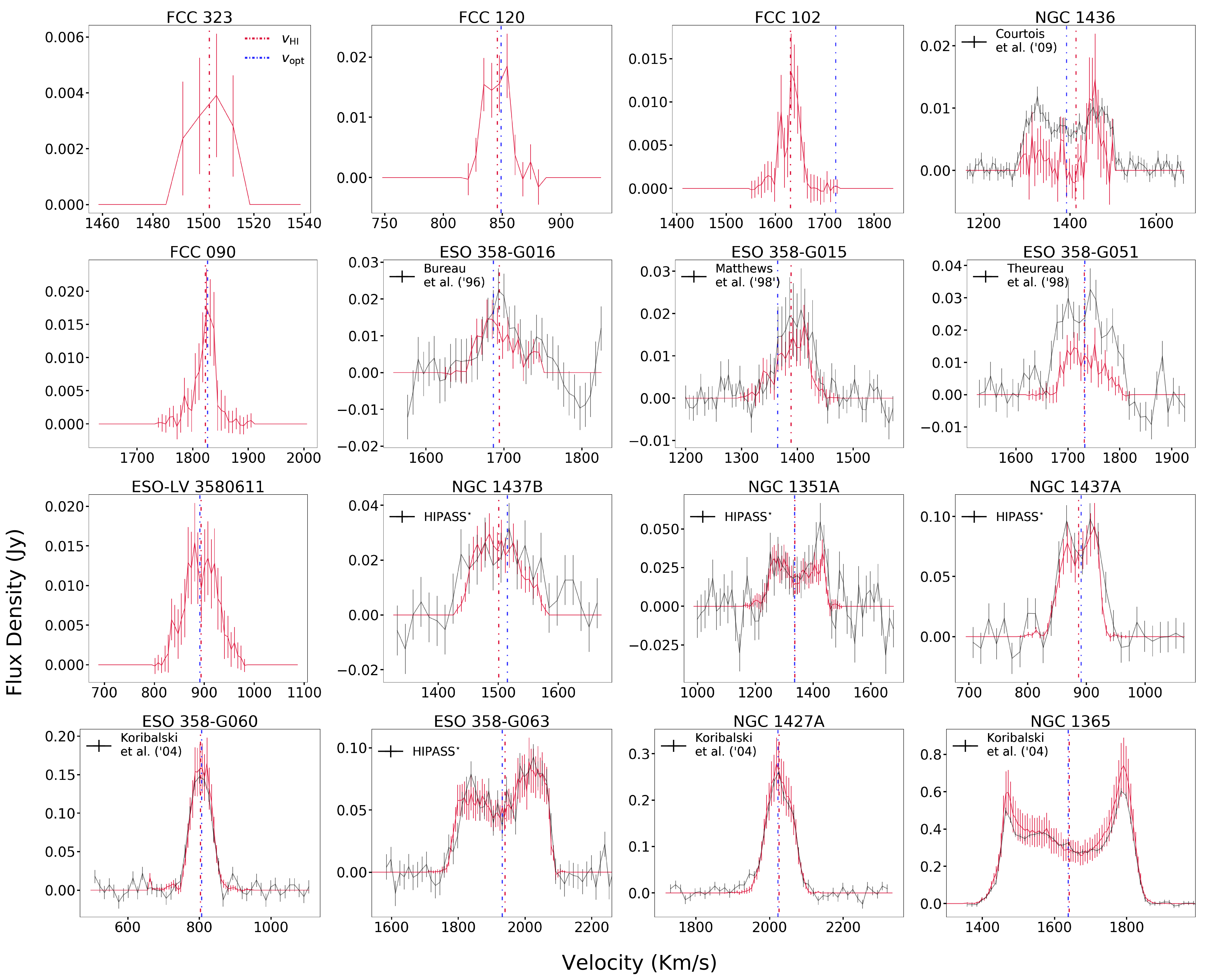}
    \caption{
     Integrated \hi{} spectra of our Fornax \hi{} detections (in red) sorted according to increasing \hi{} mass as in Fig.~\ref{fig:allmorph}. We compare our spectra to spectra from the literature (shown in black; see top-left corner for details). We also show the barycentric velocity obtained from our \hi{} spectra (vertical red line) and from optical spectra (vertical blue line; \citealp{Maddox2019}).}
    \label{fig:allprof}
    \end{figure*}{}

    \cite{Waugh2002} presented a blind survey of Fornax based on the \hi{} Parkes All Sky Survey (HIPASS $-$ \citealt{HIPASS_2001MNRAS}) data with a  \mhi{} limit of 1.4~$\times$ 10$^8$ M$_{\odot}$. 
    They detected 110 galaxies within an area of $\sim$620 deg$^2$ around the cluster. Of those, nine \hi{} detections are within $R_\mathrm{vir}$. 
    The authors confirmed a \hi{} depletion in the \hi{} detections (all late types) near the centre of the cluster, and suggested that \hi{}-rich galaxies detected in the outer parts of the cluster are infalling towards the cluster for the first time.
    \hi{} detections are arranged in a large scale sheet-like structure with a negative velocity gradient from south-east to north-west.
    
    \cite{waughphd} presented the result of the deepest blind \hi{} survey of the cluster so far: the Basketweave survey of Fornax carried out with the Parkes telescope. This survey used a new scanning technique, which improved the sampling and the noise level compared to HIPASS.
    \hi{} was detected in 53 galaxies within 100~deg$^2$ down to a detection limit of 10$^8$ M$_{\odot}$. Of those, 15 were new detections, which confirmed the results of \cite{Waugh2002}.
    In addition, the author presented higher resolution \hi{} observations (2~arcmin, 3.3 \kms{}) of 28 individual Fornax galaxies carried out with the Australia Telescope Compact Array (ATCA).
    Of those, six are within $R_\mathrm{vir}$. Within this region, ESO~LV-3580611 was marked as having an intriguing \hi{} morphology with an elongation to the north-east of the system. As suggested in \cite{Schroder2001}, also \cite{waughphd} pointed out that this galaxy may be moving towards us. 
    
    The only other Fornax galaxy within $R_\mathrm{vir}$ with resolved \hi{} imaging is NGC~1365 \citep{vanderHulst1365,Jorsater1365}. The latter study observed an elongated \hi{} distribution to the west of the system. They suggested that it may be caused by the interaction with the ICM.
    
    All previous blind \hi{} surveys of Fornax were carried out with the Parkes telescope. Their angular resolution of 15 arcmin was insufficient to study the \hi{} morphology of the detected galaxies, which is a powerful tracer of environmental effects. Better angular resolution was achieved with the interferometric observation of a few selected galaxies in Fornax (see references above), but those observations covered only a small portion of the cluster volume. Here we present the first, blind, interferometric \hi{} survey of the Fornax cluster.
    
    Our survey was carried out with the ATCA and covers an area of 15 deg$^2$ centred on NGC~1399 with a spatial and velocity resolution of 67\arcsec~$\times$~95\arcsec ($\sim6~\times~9$~kpc at a distance of 20~Mpc) and 6.6~\kms{}, respectively. The average column density sensitivity within the survey area is 2$\times$10$^{19}$~cm$^{-2}$ (3$\sigma$ over 25~\kms{}) and the \mhi{} sensitivity is 2$\times10^7$~M$_{\odot}$~(3$\sigma$ over 100~\kms{}). 
    Initially, a case study by \cite{karenNGC1427A} revealed the tidal origin of NGC~1427A using spatially resolved \hi{} images from a subregion of our ATCA mosaic.
    Here we present the results of the full survey.
    In Sect.\ref{sec:obs} we describe observations and data reduction.
    In Sect.\ref{sec:emiss} we present the \hi{} detections, their \hi{} images and spectra. We also compare their \hi{} mass, \htwo{0pt} mass and SFR relative to non-cluster control samples, and compare their spatial and velocity distribution with those of the general Fornax population. In Sect.~\ref{sec:disc} we discuss our results, which we then summarise in Sect.\ref{sec:summar}.
    We include supplementary material on each galaxy in Appendix \ref{sec:app}.
    
    \section{ATCA observations and data reduction}
    \label{sec:obs}
    Our blind Fornax survey covers an area of 15 deg$^2$ (defined at a sensitivity level 3 $\times$ higher than in the best region of the \hi{} cube), spanning from the centre of the cluster to a distance slightly further than $R_\mathrm{vir}$. The observations were carried out with the ATCA in the 750B configuration, from December 2013 to January 2014 (project code C2894)\footnote{Data available on \url{https://atoa.atnf.csiro.au/query.jsp}}. The cluster was observed for 336 hrs using 756 different pointings with a spacing of 8.6 arcmin (1/4 of the primary beam FWHM at 1.4 GHz)

    The 64~MHz bandwidth, centred at 1396~MHz, was divided into 2048 channels, providing a velocity resolution of 6.6~\kms. 
    We reduced the data using the {\tt MIRIAD} software \citep{SaultMiriad}. PKS~B1934-638 and PKS~0332-403 were chosen as the bandpass calibrator and the phase calibrator, respectively. The latter was observed at 1.5~hr intervals between on-source scans.
    We flagged strong radio frequency interference based on Stokes V visibilities. 
    After flagging and calibration we further processed a restricted frequency range 1407.2 MHz~-~1419.6 (which corresponds to the velocity range of 166~-~2783~\kms{}), which includes all spectroscopically confirmed Fornax galaxies \citep{Maddox2019}. 
    Within this range we used the {\tt UVLIN} task of {\tt MIRIAD} to fit and subtract continuum emission using 2nd-order polynomials.
    
    We obtained the dirty cube with the {\tt INVERT} task using natural weights to maximise surface-brightness sensitivity. 
    We used {\tt MOSMEM} and {\tt RESTOR} to clean and restore \hi{} emission, respectively. The restoring Gaussian PSF has a major and minor axis FWHM of 95 and 67~arcsec, respectively, and a position angle of 0.4~deg.
    The root mean square (RMS) noise level of the final cube goes down to 2.0 \mJb{} in the most sensitive region. Within the survey area the RMS noise is $\leq6.0$~\mJb{} and, on average, 2.8~\mJb{}. This corresponds to a $3\sigma$ \hi{} column density sensitivity of \nhi{}~$\sim2\times$10$^{19}$~cm$^{-2}$ assuming a line width of 25~\kms{} and a $3\sigma$ \mhi{} sensitivity of ~$\sim$2$\times$10$^7$~M$_\odot$ over a linewidth of 100~\kms{}. 
    We searched for \hi{} sources with the SoFiA source-finding package \citep{PaoloSoFiA} within the survey footprint (see the green outline in Fig.~\ref{fig:foot}). 
    By smoothing and clipping, we convolve the input cube with a set of kernels and detect emission above 3.5$\sigma$ of the local noise level of each kernel. 
    Reliable detections are identified based on the reliability algorithm presented in \citep{PaoloReliability}, which assumes that true sources have positive total flux and that the noise is symmetric around 0. For one faint source, NGC~1436, visual inspection was necessary to improve the SoFiA detection mask used to estimate galaxy parameters.
    Lastly, our final list of \hi{} detections includes a faint source which did not pass the reliability test, but which we consider a genuine detection given its spatial correspondence with the known optical source FCC~323.

    \section{\hi{} detections in the Fornax cluster}\label{sec:emiss}
    \subsection{\hi{} detection properties}\label{sec:hiproducts}

    We detect \hi{} in the 16 galaxies listed in Table~\ref{tab:mass}. Of these, three are new \hi{} detections: FCC~090, FCC~102, FCC~323. The last is the only galaxy with no previous redshift measurement.
    In Fig.~\ref{fig:foot} we show the location of our \hi{} detections on the sky and in Fig.~\ref{fig:allmorph} we show the \hi{} morphology of each galaxy. In these figures, red and white contours, respectively, represent the lowest reliable \hi{} column density contour, defined as 3 times the local RMS assuming a typical \hi{} linewidth of $\sim$25~\kms{} (see Table \ref{tab:mass} and top-right corner of each panel in Fig.~\ref{fig:allmorph}). 
    
    Fig.~\ref{fig:allprof} shows the integrated ATCA \hi{} spectra of our 16 detections.
    We calculated the error bars by summing in quadrature the statistical uncertainty $-$ derived from the local RMS (Table~\ref{tab:mass}) and the number of independent pixels detected in each channel $-$ and the flux-scale uncertainty.
    We find the latter to be $\pm20$\% based on the flux ratio calculated for a sample of selected bright point sources between our radio continuum image and the Northern VLA Sky Survey (\citealt{NVSS_1998AJ}). 

    We compare our spectra to those obtained from previous observations. In particular, we use HIPASS spectra from the BGC catalogue \citep{BaerbelBGC2004} or HIPASS data reprocessed by us. Since HIPASS data do not show any emission at the position of NGC~1436 and ESO~358-G016, we used Green Bank Telescope data (GBT - \citealt{courtoisGBT}) and Parkes data \citep{Bureau1996} as comparison, respectively. On the other hand, HIPASS spectra of ESO~358-G015 and ESO~358-G051 are noisy, so we used comparison spectra from \cite{ESO015_1998AJ....116.1169M} and \cite{ESO051_1998A&AS..130..333T}, respectively, based on Nanacy data. Furthermore the literature spectra of ESO~358-G015 and ESO~358-G016 were rebinned to our ATCA channel-width.
    For consistency, we calculated the uncertainties in the comparison spectra by combining the noise in the spectrum and the flux-scale uncertainty of each survey, except for ESO~358-G016 for which the flux-scale uncertainty was not provided. For this galaxy, the error bars are shown as the RMS of the spectrum.
    In each panel of Fig.~\ref{fig:allprof} we also show the velocity \vspe{} derived from optical spectroscopy \citep{Maddox2019} and the barycentric \hi{} velocity $v_\mathrm{HI}$ derived from our ATCA spectra. 
    
\begin{figure}[t]
        \centering
        \includegraphics[width=80mm]{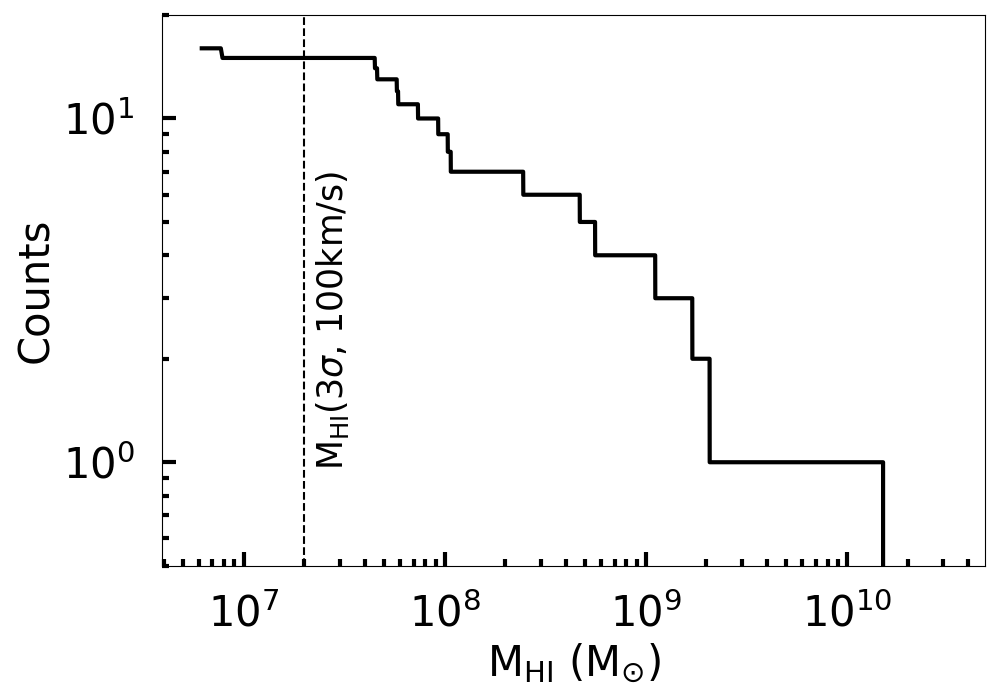} 
        \caption{Cumulative histogram of the ATCA \hi{} masses of our Fornax sample.}
        \label{fig:histo}
    \end{figure}

    We estimated the \hi{} mass (\mhi{}) of our detections from the integrated \hi{} flux using eq.50 in \cite{Meyertracing} and adopting the same distance of 20~Mpc for all galaxies \citep{2000ApJ...528..655M}. The uncertainty on \mhi{} is obtained from the error bars of the spectrum (Fig.~\ref{fig:allprof}). 
    We report \hi{} fluxes and masses in Table~\ref{tab:mass}. Our \hi{} fluxes agree with those in the literature within 1$\sigma$ for nine out of 13 galaxies, and within 2$\sigma$ for 11 of them. The two cases with a discrepancy larger than 2.5$\sigma$ are NGC 1436 and ESO~358-G051.
    The comparison spectrum for NGC~1436 comes from GBT data (\citealp{courtoisGBT}) and shows that
    we are most likely missing \hi{} flux from the blue-shifted part of the system. 
    For this galaxy, the total \hi{} flux recovered by ATCA is lower than the GBT flux by 2.5$\sigma$ (corresponding to a factor of 2.8).
    The reason of this discrepancy arises from a combination of low S/N and the presence,  in at least some of the blue-shifted channels, of artefacts in this part of the ATCA cube.
    The case of ESO~358-G051 is less clear since the ATCA cube does not show any obvious artefacts and the galaxy well detected. However the total \hi{} flux recovered by ATCA is lower than the Nancay flux by 3$\sigma$. Based on the current data it is possible that some of the emission is spread over multiple ATCA beams and therefore is too weak to be detected. Future MeerKAT data will clarify this issue \cite{2016PaoloMeerkat}.
    Finally, the \hi{} mass of FCC~323 is below the typical sensitivity of our data quoted in Sec.~\ref{sec:obs} because of the narrow linewidth as well as the low value of the local noise (Table~\ref{tab:mass}).
    Our detections cover about three of magnitude in \mhi{}, from FCC~323 (\mhi{}~=~$8\times10^6$~M$_\odot$) to NGC~1365 (\mhi{}~=~$1.5\times10^{10}$~M$_\odot$). Fig.~\ref{fig:histo} shows the cumulative histogram of the \hi{} masses of our sample. A future study will analyse the \hi{} mass function inferred from our data. 
    
    {}
    
        \begin{figure}[t]
        \centering
        \includegraphics[width=80mm]{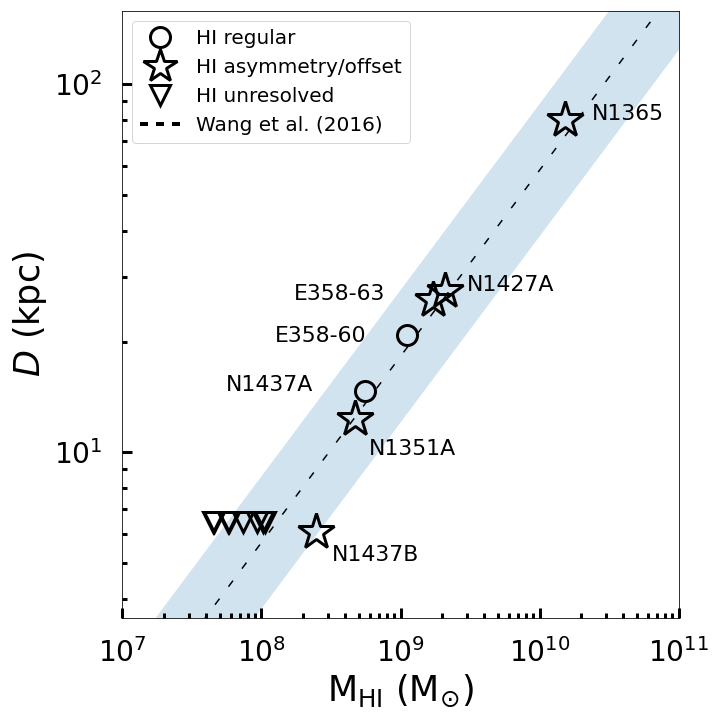}
        \caption{\hi{} disc size as a function of ATCA \mhi{} for our resolved \hi{} detections. Dashed line and the blue shaded area show the scaling relation and 3$\sigma$ scatter respectively in \citep{wang2016_size}.}
        \label{fig:HIsizerel}
    \end{figure}{}
    
    Due to the improved resolution of ATCA over a single dish, we detected, in half of the sample, a variety of \hi{} morphologies (see Fig.~\ref{fig:allmorph}) including offsets between optical and \hi{} centres, truncated discs, asymmetries and \hi{} tails, which we describe in this section and in Appendix~\ref{sec:app}, following the same order in which galaxies are shown in Fig.~\ref{fig:allmorph} (from the lowest to the highest \hi{} mass): the \hi{} distribution in FCC~102 is offset with respect to the optical centre towards the north. We also notice a large difference between \vspe{} and \vbary{}, $\sim$100 \kms{}. However, this is consistent with the large uncertainty on \vspe{} given that the latter was measured from absorption lines for this galaxy (Natasha Maddox, priv. comm.); the \hi{} peak of FCC~090 corresponds to the optical centre but the \hi{} distribution has an elongation to the south; \hi{} in ESO-LV~3580611 (FCC~306) is more extended to the north although there is no offset between optical and \hi{} centre \citep{waughphd};
    the \hi{} morphology of NGC~1437B (FCC~308) is asymmetric and more extended to the south;
    \hi{} in NGC~1351A (FCC~067) shows an elongation towards the south;
    the \hi{} of ESO~358-G063 (FCC~312) is more extended to the east side of the disc and the \hi{} contours appear to be compressed on the west side;
    In NGC~1427A (FCC~235), we detected a \hi{} tail which points to the south-east, away from the cluster centre, consistent with the tidal origin of this galaxy discussed in \cite{karenNGC1427A};
    the \hi{} distribution in NGC~1365 (FCC~121) is extended to the north and appears to be compressed in the south-west part of the disc (see also \citealt{Jorsater1365}).     
    We mark these \hi{} disturbed galaxies with star markers in all subsequent figures.

    The remaining half of the galaxies are either unresolved (or nearly so) and centred on the stellar body, or do not show noticeable asymmetries.
    One of them hosts a \hi{} disc unusually truncated within the stellar disc, NGC~1436. This case will be discussed in detail in later sections.
    \begin{figure}[t]
        \centering
        \includegraphics[width=80mm,align=t]{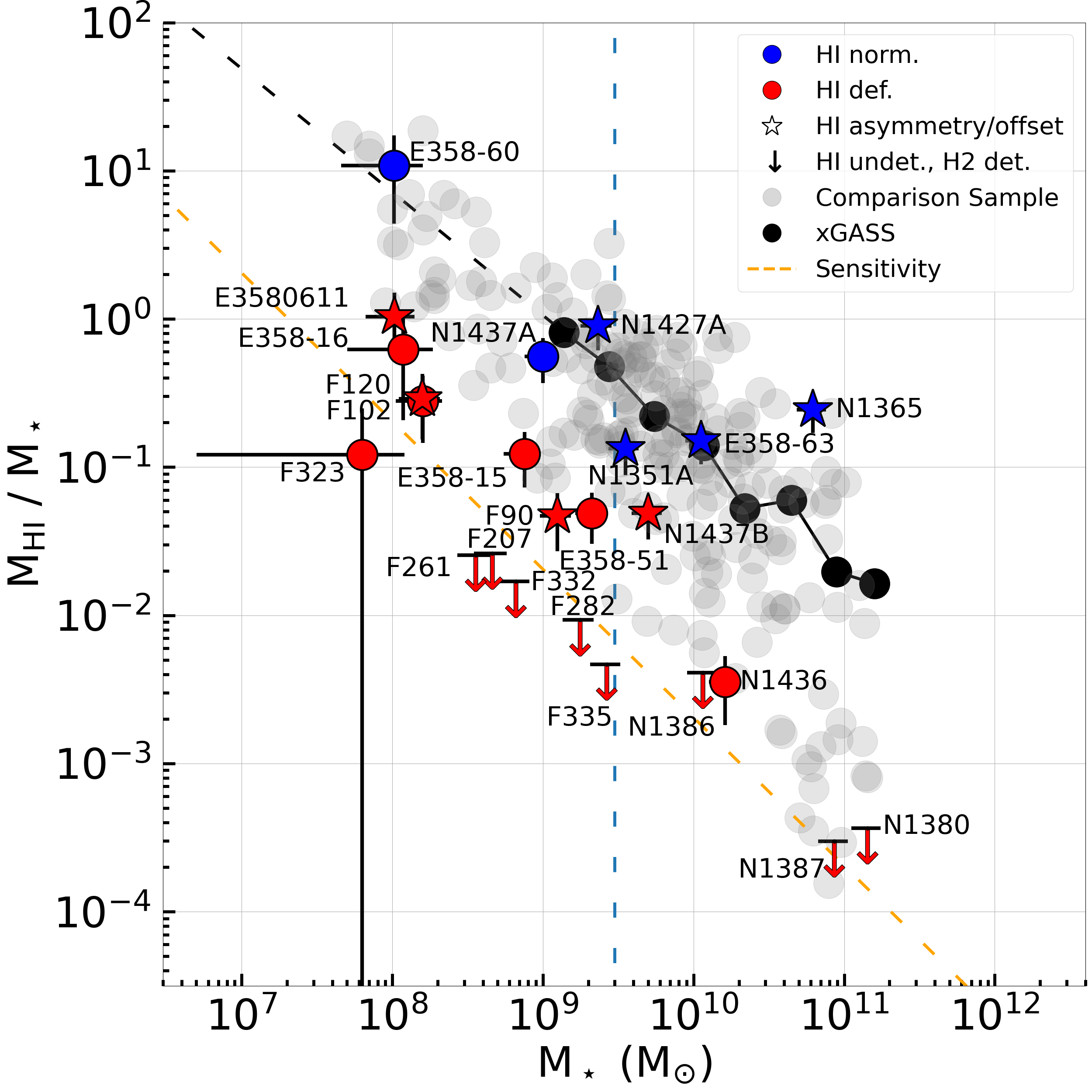}
        \captionof{figure}{\mhi{} to \mst{} ratio as a function of \mst{}. Fornax galaxies (red+blue markers) are compared with non-cluster galaxies from VSG+HRS (grey circles). Red and blue colours show Fornax \hi{} deficient and normal galaxies, respectively. We show Fornax galaxies with a distorted \hi{} morphology with star-shaped markers (see Sect-~\ref{sec:emiss}). We show with a black solid line the xGASS scaling relation. The black dashed line is the linear extrapolation of this trend for \mst{}~$<$~1.4~$\times$~10$^9$M$_\odot$. The orange dashed line shows the ATCA average sensitivity evaluated as 3 $\times$ 2.8 \mJb{} with a linewidth of 100 \kms{}. The vertical dashed line at 3~$\times$~10$^9$~M$_\odot$ separates low- from high-mass galaxies.}
        \label{fig:himfraction}
    \end{figure}{}

    A final case worth commenting on is FCC~120. This galaxy has been detected with Parkes and the spectrum in \citet{Schroder2001} shows an evident double horn profile that is $\sim$100\kms{} wide, while we detect a single-peak profile (within the uncertainties - see Fig.~\ref{fig:allprof}).
    Careful visual inspection of our \hi{} cube did not reveal any \hi{} emission missing from our SoFiA detection mask. The regular \hi{} morphology of this galaxy in Fig.~\ref{fig:allmorph} and the good agreement between \vbary{} and \vspe{} in Fig.~\ref{fig:allprof} suggest that our \hi{} characterisation of this galaxy is correct. Future, deeper data from the Meerkat Fornax survey may confirm this (\citealp{2016PaoloMeerkat}).
    
    We estimated  the \hi{} size for all our resolved \hi{} detections using the method of \citealt{wang2016_size} from the \hi{} intensity maps of our galaxies. We considered a galaxy resolved if its surface brightness profile deviates from the shape of the point-spread function. We measured the HI diameter where the surface density is 1~M$_\odot$~pc$^{-2}$ and is then deconvolved with the \hi{} beam.
    In Fig.~\ref{fig:HIsizerel} we show that both disturbed and regular galaxies follow the \hi{}~size-mass scaling relation of \citet{wang2016_size} within the 3$\sigma$ scatter. This can be understood as asymmetric \hi{} features usually have a low surface brightness and do not significantly contribute to the total \mhi{} of galaxies. In Fig.\ref{fig:HIsizerel} we also show the \hi{} unresolved galaxies. The upper limit on their size were set equal to the ATCA beam minor axis.
    
        \begin{figure}[t]
        \centering
        \includegraphics[width=80mm]{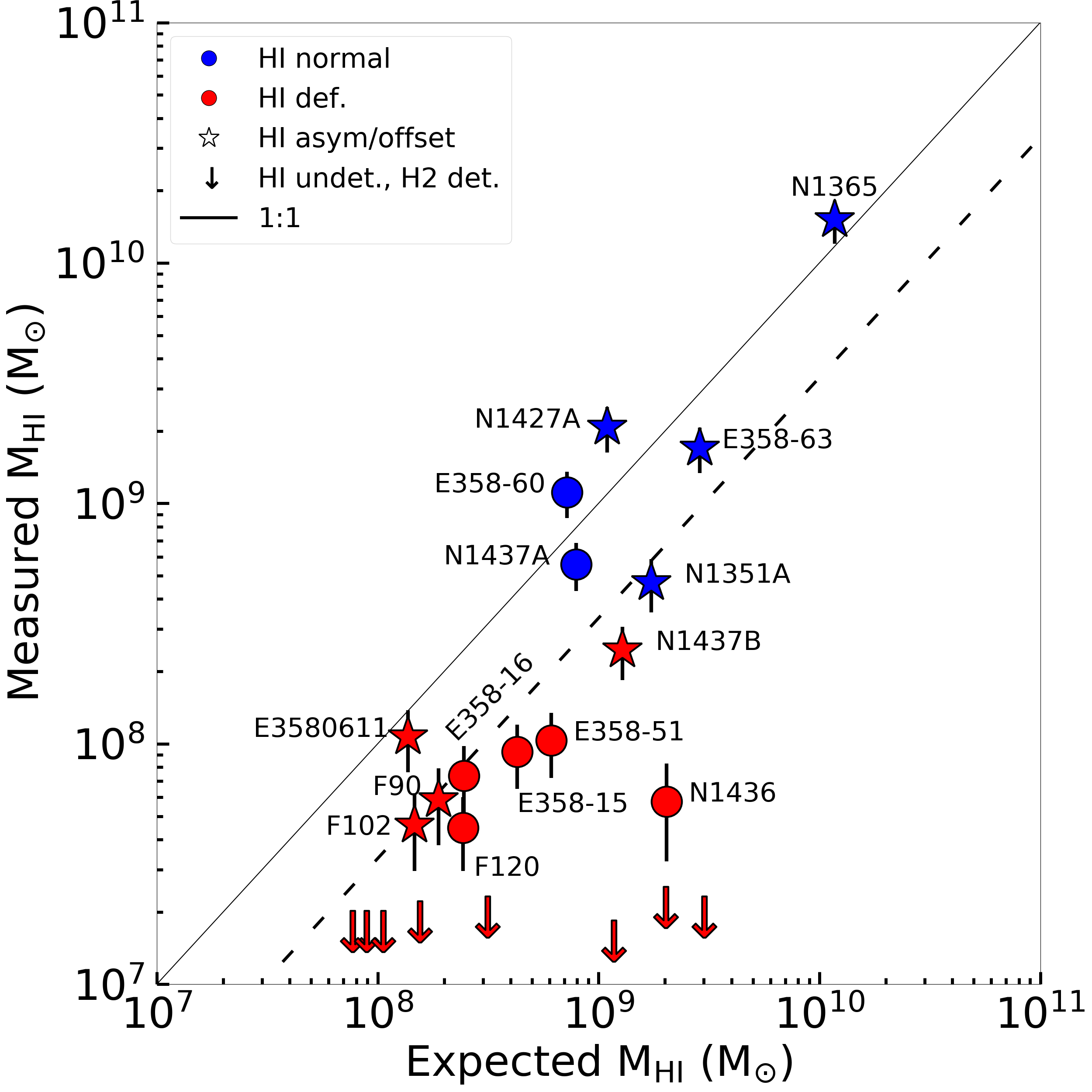}
        \caption{Expected \mhi{} versus measured \mhi{}. The former were evaluated with the \citeauthor{H&G84}'s (\citeyear{H&G84}) method with the coefficient summarised by \cite{BoselliGavazzi2009}. The latter are the ATCA \mhi{} calculated as described in Sect.\ref{sec:hiproducts}. The dashed line, a factor of 3 below the 1:1 relation, shows the typical threshold below which galaxies are considered deficient.} 
        \label{fig:HG}
    \end{figure}{}
    
    \subsection{\mhi{} to \mst{} ratio}
    \label{subse:himfrac}
    Given the abundant (albeit subtle) \hi{} asymmetries and offsets in our Fornax sample, which might be tracing environmental interactions within or on their way to the Fornax cluster (whether with other galaxies, the large-scale potential or the intergalactic medium), we study the amount of \hi{} in these galaxies in search of signs of \hi{} depletion.
    In order to do so, we evaluated the ratio between \mhi{} and stellar mass (\mst{}) for each galaxy in our sample.
    The \mst{} values are derived using the WISE W1 (3.4~$\mu$m) in$-$band luminosity, W1$-$W2 colour and the prescription given by \cite{2014wise}, with the custom photometry further defined in \cite{2019tom}, and assuming a common distance of 20~Mpc for all galaxies. We show the distribution of our sample on the \mhi{}/\mst{}$-$vs$-$\mst{} plane in Fig.~\ref{fig:himfraction}.
    
    We compare galaxies in Fornax with a sample consisting of void galaxies from the Void Galaxy Survey (VGS - \citealt{VoidKreckel2012}) and field galaxies from the Herschel Reference Survey (HRS - \citealt{HRSboselli2014}). 
    We also used the xGASS \mst{}-\mhi{}/\mst{} scaling relation, which shows the weighted median of $\log_{10}$(\mhi{}/\mst{}) as a function of $M_{\star}$. This was obtained from 1177 galaxies selected only by stellar mass (\mst{}~=~10$^9$–10$^{11.5}$~M$_{\odot}$) and redshift (0.01~$<$~z~$<$~0.05; \citealt{BarbaraxGASS}). In Fig.~\ref{fig:himfraction} we extrapolate the xGASS trend down to 10$^7$ M$_\odot$ (black dashed line) and show that it is consistent with the non-cluster sample of galaxies.
    
    Our sample of Fornax \hi{} detections appears to be systematically
    offset with respect to our VGS+HRS comparison sample
    and to the xGASS scaling relation. A two-sample Kolmogorov–Smirnov test
    on the distribution of off sets from the xGASS scaling relation rejects the null hypothesis
    that Fornax and VGS+HRS galaxies are drawn from
    the same parent sample (p-value = 0.004, KS statistic~=~0.43).
    Indeed, ten out of 16 Fornax
    galaxies are below or at the lower edge of the distribution of non-cluster
    galaxies. Their offset from the xGASS scaling relation is
    larger than the RMS deviation of VGS+HRS galaxies from it,
    thus indicating \hi{} deficiency. For illustrative purposes we henceforth
    label these HI deficient galaxies in red colours in Fig. 6 and
    in all other upcoming figures in this paper.
    
    \hi{} deficiencies measured from plots like our Fig.~\ref{fig:himfraction} should be taken with caution
    because of the large \mhi{} scatter at at fixed \mst{} (e.g. \citealt{Maddox2015}).
    This scatter is to first order driven by Hubble type or, equivalently, by galaxy properties that correlate with Hubble type, such as the star formation rate (SFR). For this reason, in Sect.~\ref{subsec:comph2} we further analyse how the \hi{} deficiency at fixed \mst{}
    relates to the \htwo{0pt} and SFR of these galaxies. Furthermore, in the present section we estimate \hi{} deficiencies as proposed by \cite{H&G84} and recently revisited by \cite{2018A&A...609A..17J}, where the measured \mhi{} is compared with an expected \mhi{} calculated based on galaxies’ optical size and Hubble type. For this purpose, we use the coefficients summarised by \cite{BoselliGavazzi2009} and originally given in \cite{H&G84}, \cite{SolanesHImassfunc}, \cite{BoselliGavazzi2009}, and
    adopt the optical sizes and Hubble types listed in Table~\ref{tab:mass} for our galaxies.
    Fig.~\ref{fig:HG} shows the comparison between the expected and the measured \mhi{}. The dashed line, a factor of 3 below the 1:1 relation, is the typical threshold below which galaxies are considered \hi{} deficient using the \citeauthor{H&G84}'s (\citeyear{H&G84}) method (e.g. \citealt{LucaVirgo2011}). Also in this case, Fornax galaxies appear offset towards lower \hi{} masses and there is a good match between galaxies labelled as \hi{} deficient based on our Fig.~\ref{fig:himfraction} and those below the dashed line in Fig.~\ref{fig:HG}.
    
    From these figures, we see that not all galaxies with a disturbed \hi{} morphology are \hi{} deficient. 
    Thus, \hi{} morphological disturbances, whatever their exact nature (e.g. tidal or hydrodynamical, which is difficult to establish with the current data), allow us to identify cases of environmental interactions before a significant fraction of the cold interstellar medium is removed. Combining \hi{} morphological information and \mhi{}-\mst{} ratio may reveal likely new members of the cluster (we come back to this point in Sect.\ref{sec:disc}). 

    Fornax galaxies with a disturbed \hi{} morphology cover the full \mst{} range.
    However, we measure a stronger \hi{} depletion for low mass galaxies: the average offset from the xGASS scaling relation in Fig.~\ref{fig:himfraction} is $-0.86$~dex and $-0.33$~dex for \mst{} below and above 3$\times$10$^9$, respectively. We show the threshold of 3$\times$10$^9$~M$_\odot$ with a vertical dashed blue line in Fig.~\ref{fig:himfraction}.
    
    We already mentioned the problematic detection of NGC~1436. Although we are probably missing some flux, 
    it remains a deficient galaxy even if we estimate the \hi{} mass from the GBT 
    flux (Table \ref{tab:mass}) with a log$_{10}$(\mhi{}/\mst{})~=~$-2.0$ and an offset from the xGASS scaling relation of $-0.93$~dex.
    
    Fig.~\ref{fig:himfraction} shows also eight galaxies where we did not detect \hi{} emission but \htwo{0pt} was detected with with the Atacama Large Millimeter/submillimeter Array (ALMA) \citep{zabel}. For these galaxies we calculated the \mhi{} upper limit as $3\times$ the local noise (Table \ref{tab:mass}) of the cube and assuming the CO line width of these galaxies estimated by the PV diagrams in \cite{zabel}. Some of these galaxies appear to have too little \hi{} given the molecular gas content and star formation rate, as we discuss in following sections. We show the optical morphology of these galaxies in Appendix~\ref{sec:app2}.

    \begin{figure}[t]
        \centering
        \includegraphics[width=80mm, align=t]{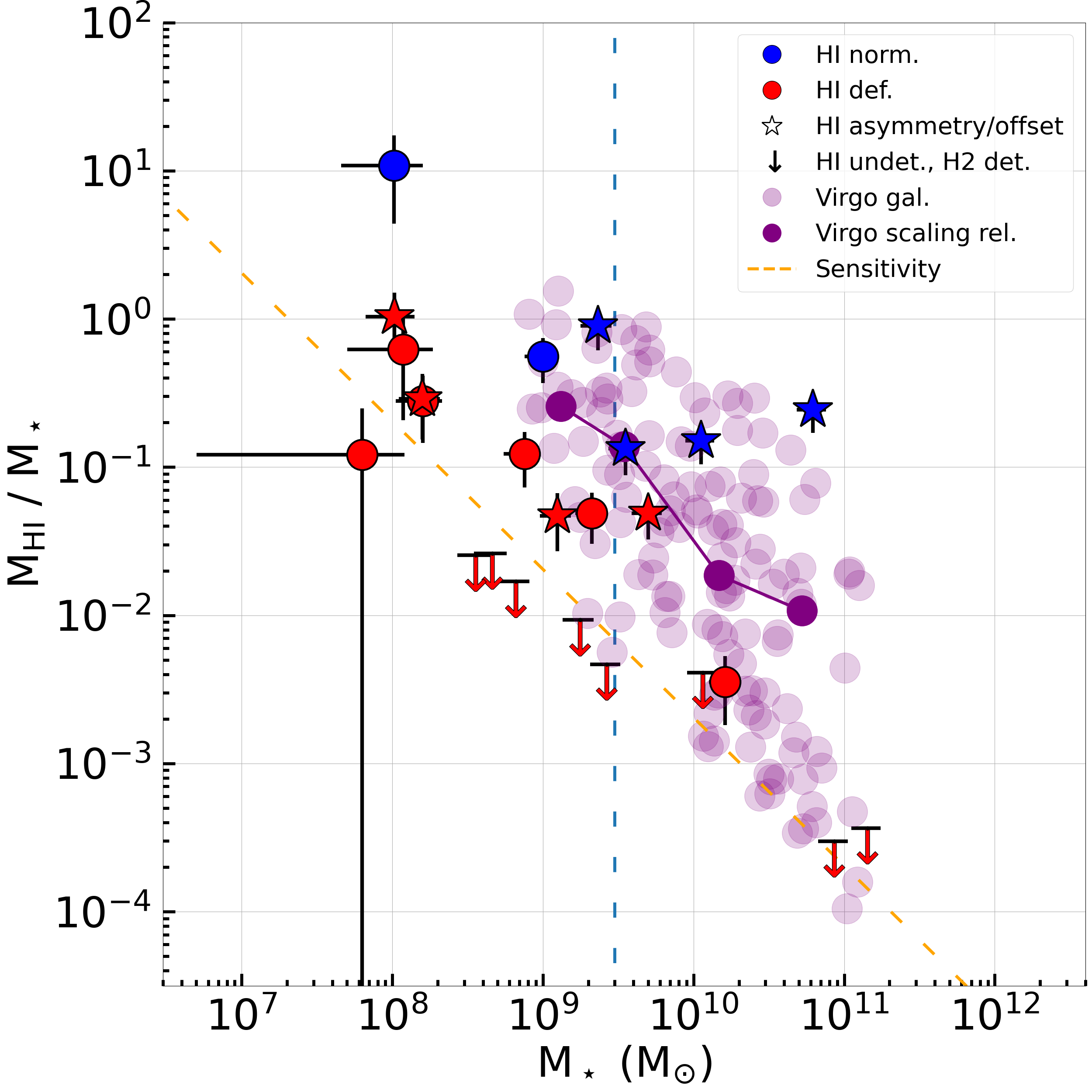}
        \captionof{figure}{We compare Fornax galaxies (same colour coding of Fig.\ref{fig:himfraction}) with Virgo cluster galaxies from HRS (light purple circles). Dark purple circles show the average scaling relation obtained from Virgo cluster galaxies \citep{LucaVirgo2011}}
        \label{fig:hivirgo}
    \end{figure}{}
    
    Finally, Fig. \ref{fig:hivirgo} shows the comparison between Fornax and Virgo galaxies (from HRS - \citealt{HRSboselli2014}) belonging to Virgo clouds A, B, N, E and S (as defined by \citealt{1999MNRAS.304..595G}) with 10$^9\lesssim$\mst{}$\lesssim$10$^{11}$~M$_\odot$. Here we show the Virgo cluster galaxies and the average scaling relation obtained from them in the same \mst{} range \citep{LucaVirgo2011}. Since Virgo is populated by \hi{} poor galaxies with respect to field galaxies \citep{1973MNRAS.165..231D, 1980A&A....83...38C, 1994AJ....107.1003C, HughesCortese2009_environm, ChungVIVA2009}, this scaling relation is shifted towards lower gas fractions with respect to xGASS (Fig.~\ref{fig:himfraction}). 
    Although Fornax and Virgo galaxies experience different cluster environments, the distribution of Fornax galaxies cover the whole range of \mhi{}/\mst{} of Virgo galaxies, reaching the same level of \hi{} deficiency.

    \subsection{Linking \hi{} properties with \htwo{0pt} and SFR}
    \label{subsec:comph2}
    
    \begin{figure}[t]
        \centering
        \includegraphics[width=80mm]{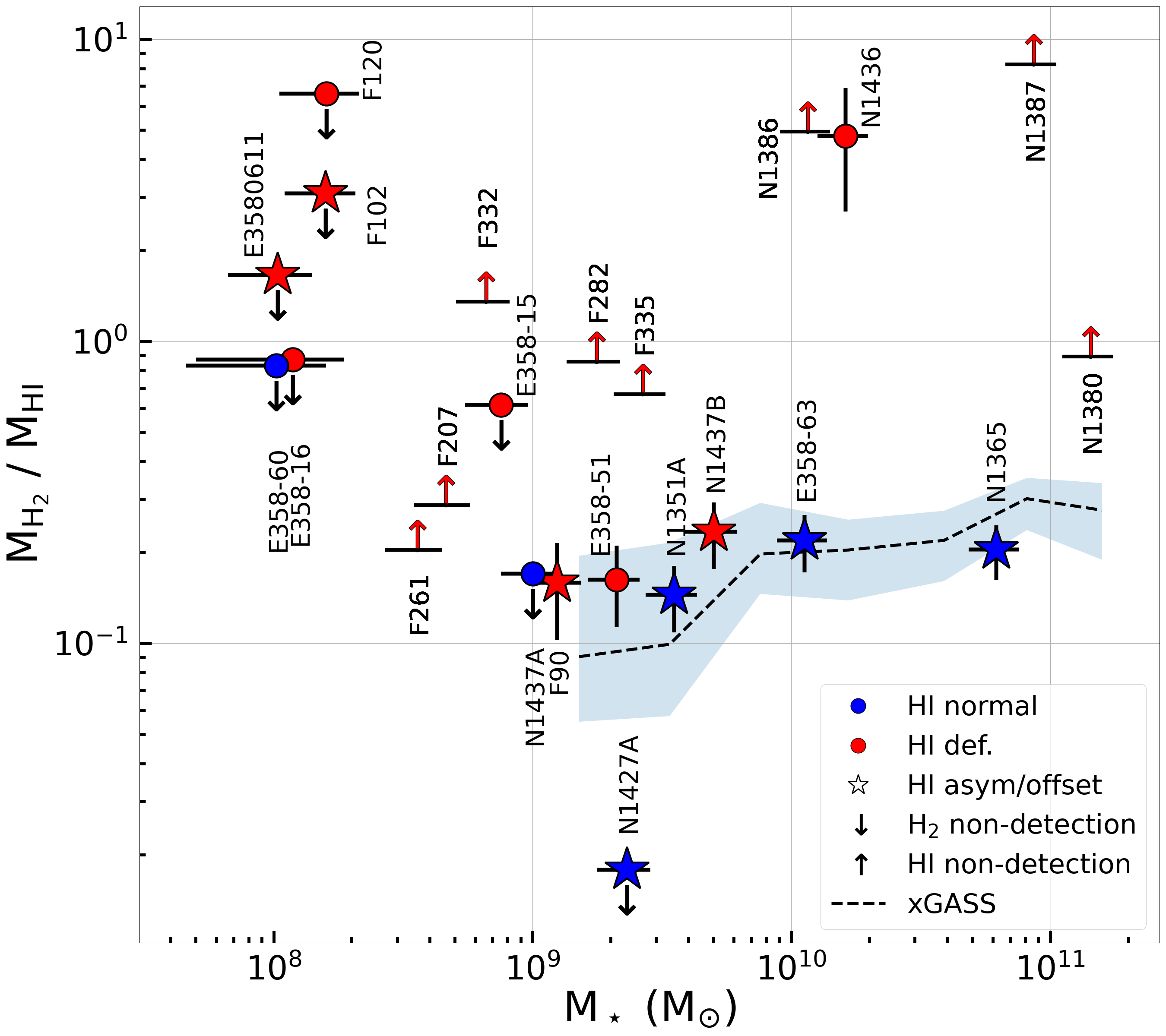}
        \caption{\mhtwo{}$/$\mhi{} as a function of \mst{}. We use the same colour coding as Fig.~\ref{fig:himfraction}. Blue shadow shows $1\times\sigma$ scatter from the xGASS weighted average of $\log_{10}$(\mhtwo{}/\mhi{}).
        We show upper limits with downward arrows. We show lower limits with upward arrows.}
        \label{fig:hionh2}
    \end{figure}{}
     \begin{figure}[t]
        \centering
        \includegraphics[width=80mm]{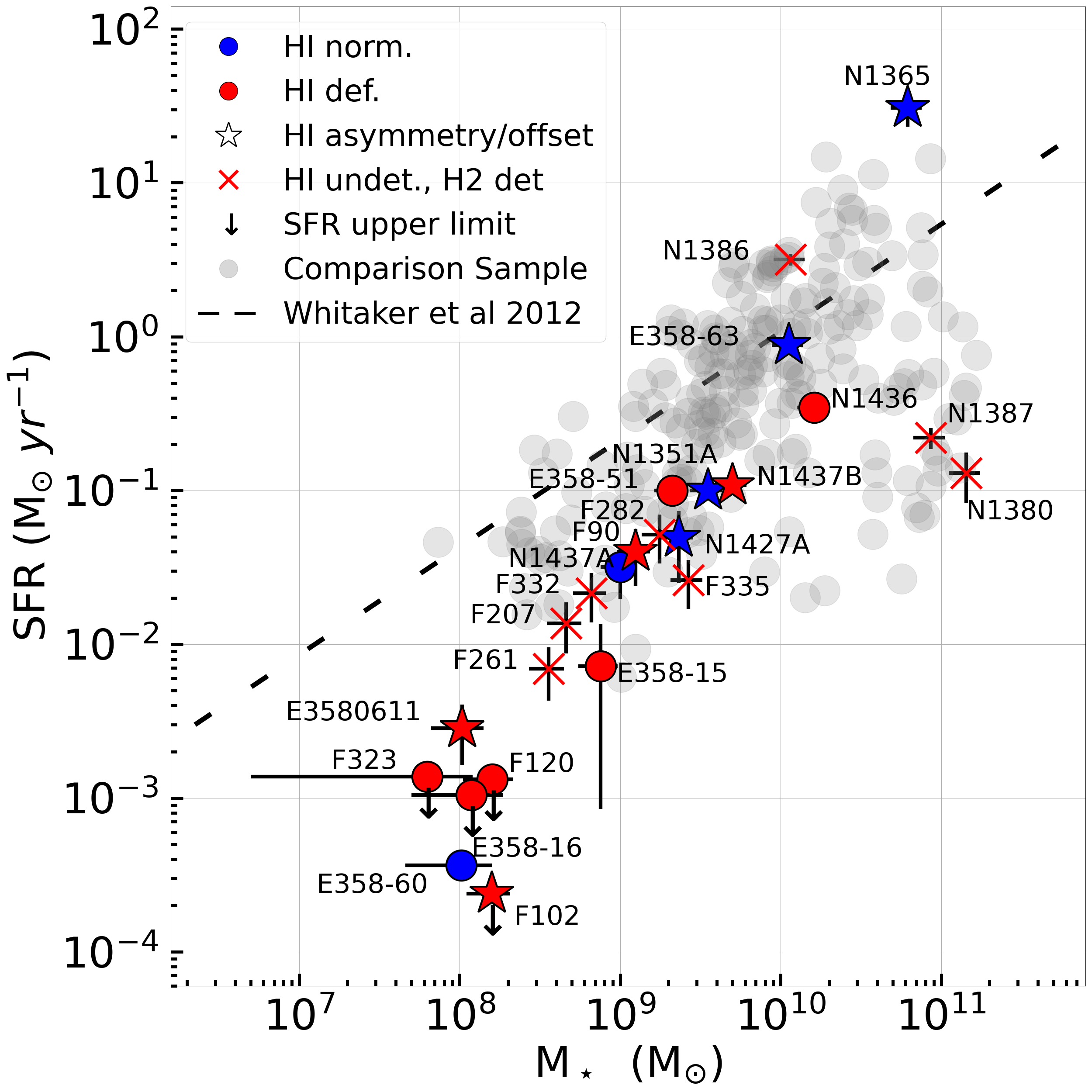}
        \caption{SFR as a function of \mst{}. We use the same colour coding as Fig.~\ref{fig:himfraction}. 
        We show upper limits in SFR with downward arrows. The dashed black line represents the SFR scaling relation in \citep{Whitaker2012ApJ}.}
        \label{fig:sfr}
    \end{figure}{}   
    
    The ratio between molecular and atomic gas mass (\mhtwo{}$/$\mhi{}) can be useful to identify anomalous galaxies where, for example, only the atomic phase is affected by the environment or \hi{} is not efficiently converted to \htwo{0pt}. 
    We thus compared the atomic and molecular gas reservoirs of our detections (see Fig.~\ref{fig:hionh2}). For this purpose, we used \htwo{0pt} masses from \cite{zabel} for all our \hi{} detections except for FCC 323 (no molecular gas data  available). We also include in this analysis the eight galaxies detected with ALMA \citep{zabel} that are not detected in \hi{} (upper limits in Fig.~\ref{fig:himfraction} and Fig.~\ref{fig:hivirgo}; see Sec.~\ref{subse:himfrac}). We also scaled the molecular upper limit from \cite{zabel} to be consistent with a line width of 100~\kms{}.
    As a comparison, we used the xGASS \mst{}-\mhtwo{}$/$\mhi{} scaling relation in \cite{BarbaraxGASS}, which describes the typical \mhtwo{}$/$\mhi{} ratio as a function of \mst{} at $z$~$=$~0.

    For \mst{}$>$10$^9$~M$_{\odot}$, 55$\%$ of all \hi{} detected galaxies are also \htwo{0pt} detected, while this fraction drops to zero for \mst{}$<$10$^9$~M$_{\odot}$ (likely because the lower metallicity of these objects makes CO progressively harder to detect). In the rest of this section we therefore focus on the higher \mst{} range.
    Here, we see that about half of the galaxies are consistent with the xGASS sample, while most of the remaining galaxies are above the xGASS scaling (see Fig.~\ref{fig:hionh2}).
    In particular, galaxies with a distorted \hi{} morphology are compatible with the xGASS trend with the exception of NGC~1427A whose lack of molecular gas is puzzling (see \citealp{zabel}). Although this galaxy has a normal (and large) \hi{} mass for its stellar mass, no molecular gas was detected with ALMA.
    The agreement with the scaling relation holds also for NGC~1437B, which is the only \hi{} deficient galaxy with a distorted morphology in this range of \mst{}. 

    The detection with the highest \mhtwo{}/\mhi{} ratio is NGC~1436. Although we might be missing \hi{} flux (Sect.~\ref{subse:himfrac}), this mass ratio remains high even if we use the \mhi{} value estimated from the GBT flux (with a \mhtwo{}/\mhi{}~=~1.7). 
    Other three galaxies, NGC~1386, NGC~1387 and NGC~1380, may be even more peculiar as the lower limit on their \mhtwo{}/\mhi{} ratio is already an order of magnitude above the xGASS scaling relation. 
    High \mhtwo{}/\mhi{} ratios have also been measured in Virgo galaxies \citep{2016MNRAS.459.3574C}.

    Given the broad distribution of \mhtwo{}/\mhi{} ratio in Fornax, we further investigate whether their star formation rate follows standard scaling with \mst{} and \mhi{}. In particular, we are interested in understanding whether the general offset of Fornax galaxies towards low \mhi{}/\mst{} ratios in Fig. \ref{fig:himfraction} is associated with a decrease of SFR at fixed \mst{}.
    
        \begin{figure}[t]
        \centering
        \includegraphics[width=80mm]{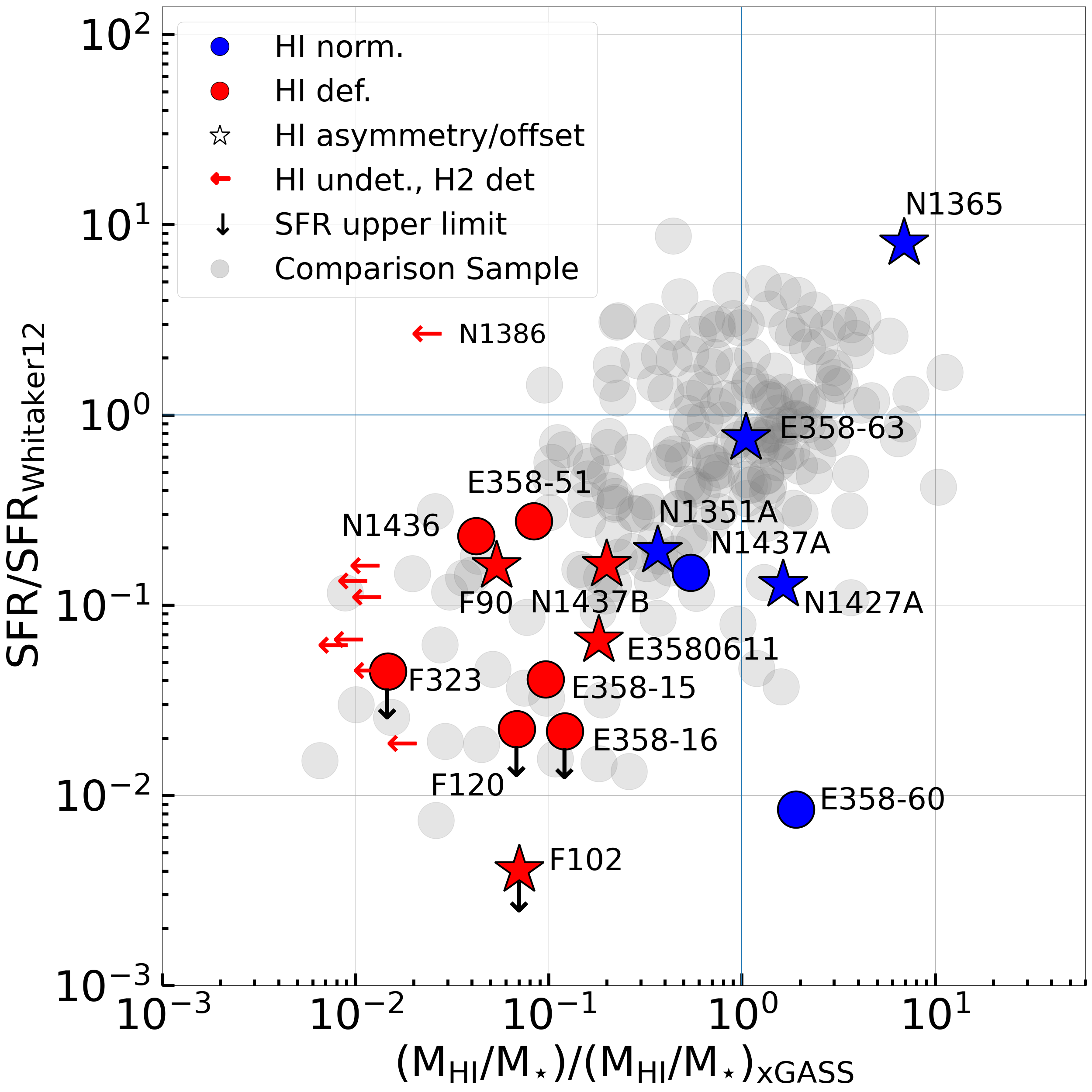}
        \caption{
        SFR deviation from the \cite{Whitaker2012ApJ} scaling relation (Fig. \ref{fig:sfr}) plotted against the \mhi{}/\mst{} deviation from the xGASS scaling relation (Fig.~\ref{fig:himfraction}).
        We use the same colour coding as Fig.~\ref{fig:himfraction}.
        We show upper limits in \mhi{} with leftward arrows. The horizontal blue line represents no deviation from the SFR scaling relation, while the vertical blue line represents no deviation from the \mst{}-\mhi{}/\mst{} scaling relation.}
        \label{fig:deltasfr}
    \end{figure}{}   
    
    We start by comparing Fornax galaxies with the same sample of non-cluster galaxies used in Sect.~\ref{subse:himfrac}. 
    SFR of both samples of Fornax galaxies and non-cluster galaxies are evaluated using eq.~2 in \citet{boquien2016A}.
    We set the scaling coefficient for the infrared 24~$\mu$m band (W4) equal to 6.17. We adopted the calibration factor for near UV 231~nm band (NUV) to be log$_{10}$C~=~-43.17 \citep{kenni2012}.
    NUV fluxes come from GCAT \citep{GCAT2012} and IRSA catalogues \citep{Leroy2019ApJS}. Since the latter catalogue is more reliable for galaxies at $z$~=~0, we used it to get the NUV flux from Fornax galaxies.
    W4 was obtained using custom software optimised for performing aperture photometry on resolved galaxies (\citealt{Tom2013AJ}; \citealt{cluver2014ApJ}; \citealt{2019tom}).
    Then, we removed the contribution from the evolved stellar populations using the method of Helou et al. (2004) which consists of scaling and subtracting the W1 light (a proxy of evolved stellar population) from W4  as described in \citet{cluver2017ApJ}.
    For a fraction of galaxies, NUV and W4 fluxes were not available (3/212 non-cluster galaxies; 11/24 Fornax galaxies). In these cases, we calculate the SFR from the W3 (12$\mu$m) luminosity -- with stellar emission subtracted (as for W4)-- and the TIR-to-MIR relation using eq.~4 in \citet{cluver2017ApJ}, where L$_{12}$~$\mu$m is the continuum-subtracted spectral ($\nu~ \times$~L$_\nu$) luminosity. Our conclusions below do not change if we calculate SFR from WISE data alone.

    Fig.~\ref{fig:sfr} shows that the majority of the Fornax galaxies have a SFR below the values predicted by the scaling relation of \cite{Whitaker2012ApJ}.
    Furthermore, the difference between the SFR of Fornax galaxies and that predicted from \citet{Whitaker2012ApJ} increases towards lower \mst{}. That is, the SFR-\mst{} relation in Fornax is steeper than that of non-cluster galaxies. This might indicate a stronger SFR decrease in low-mass galaxies, similar to what observed for their \hi{} reservoirs (as discussed in Sect.~\ref{subse:himfrac}) and their \htwo{0pt} reservoirs \citep{zabel}. Also when compared to the HRS+VGS comparison sample, Fornax galaxies appear to be offset towards lower SFR values, mirroring the results in Fig. \ref{fig:himfraction}.
    
    Among the four galaxies with the highest \mhtwo{}/\mhi{} ratio in Fig.~\ref{fig:hionh2}, NGC~1386 is the only one with a higher SFR than expected. This is likely due to the presence of an AGN \citep{2017MNRAS.470.2845R} which affects the SFR measurement.
    The other two \hi{} undetected galaxies, NGC~1380 and NGC~1387, reside at the lower edge of the comparison sample. We notice that NGC~1380 is also highly \htwo{0pt} deficient in \citet{zabel}.
    
    \begin{figure}[t]
        \centering
        \includegraphics[width=80mm]{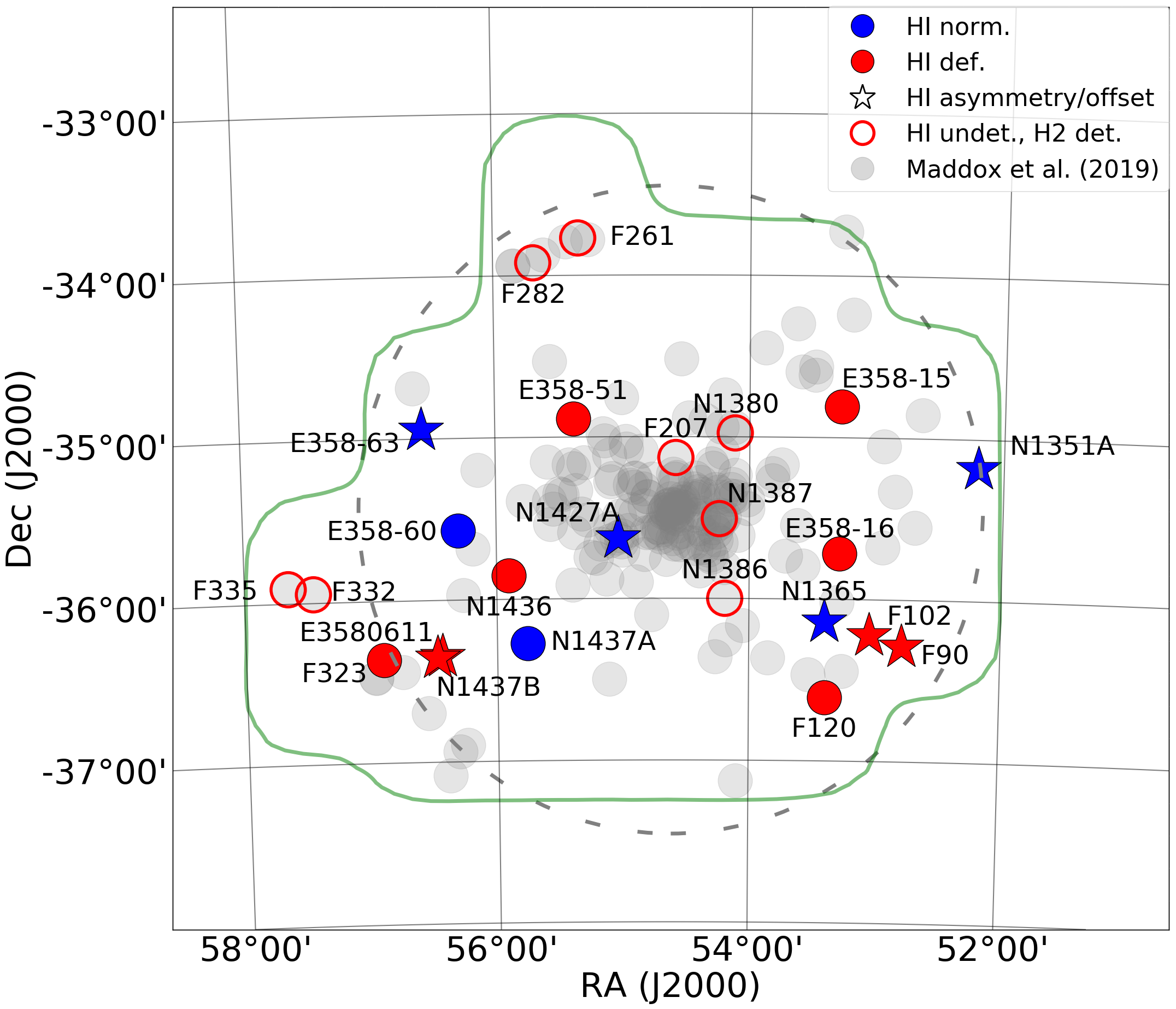}
        \caption{Distribution of our Fornax \hi{} detections on the sky (blue and red markers) compared to that of all Fornax galaxies in our footprint (grey circles; \citealp{Maddox2019}). The green and gray contours are the same as Fig.~\ref{fig:foot}.}
        \label{fig:distrib}
    \end{figure}{}
    
    In Fig.~\ref{fig:deltasfr} we investigate whether, in Fornax, the low SFR (at fixed \mst{}; Fig.~\ref{fig:sfr}) can be related to the \hi{} deficiency (Fig. \ref{fig:himfraction}). This figure plots the SFR and \mhi{}/\mst{} deviations from the respective scaling relations (Fig.~\ref{fig:sfr} and Fig.~\ref{fig:himfraction}) against one another. We find that Fornax galaxies mostly populate the area of the plot of \hi{} deficient galaxies with low SFR, following the trend of non-cluster galaxies in the comparison sample. That is, in Fornax, a decrease in \mhi{} is accompanied by a decrease in SFR just like in non-cluster galaxies. In Sec.~\ref{sec:disc} we present an interpretation of this result.

    On top of this general trend we see a few outliers.
    The most evident ones are ESO~358-G060 and the \hi{} undetected NGC~1386. While the high SFR in NGC~1386 might be an artefact caused by the presence of an AGN, the case of ESO~358-G060 is more complicated. We carefully inspected this galaxy's WISE images and no SFR from W3/W4 with old stellar population subtraction was detectable. Therefore, we calculated its SFR the from NUV flux alone using eq.~12 in \citet{kenni2012}. Although ESO~358-G060 is \hi{} rich, its SFR is lower than that any galaxy of the comparison sample with a similar \mst{}. 
    It is worth noting the cases of NGC~1436  which lies at the upper edge of the comparison sample of \hi{} deficient galaxy (bottom-left quadrant of Fig.~\ref{fig:deltasfr}) and NGC~1427A which shows a low SFR although its large \hi{} reservoir.
    All these cases will be discussed in Sect.~\ref{sec:disc}.

    \subsection{\hi{} properties as a function of 3D location in Fornax cluster}
    \label{subsec:distr}
    
    In Fig.~\ref{fig:distrib}, we compare the 2D distribution of our Fornax \hi{} detections with that of all spectroscopic Fornax members \citep{Maddox2019} included within our survey footprint. 
    Keeping in mind that the area of the sky that we observed is not symmetric, most of the galaxies in our sample are located south of the centre of the cluster, while only four out of 16 detections are located north of it. 

    \begin{figure}[t]
        \centering
        \includegraphics[width=80mm]{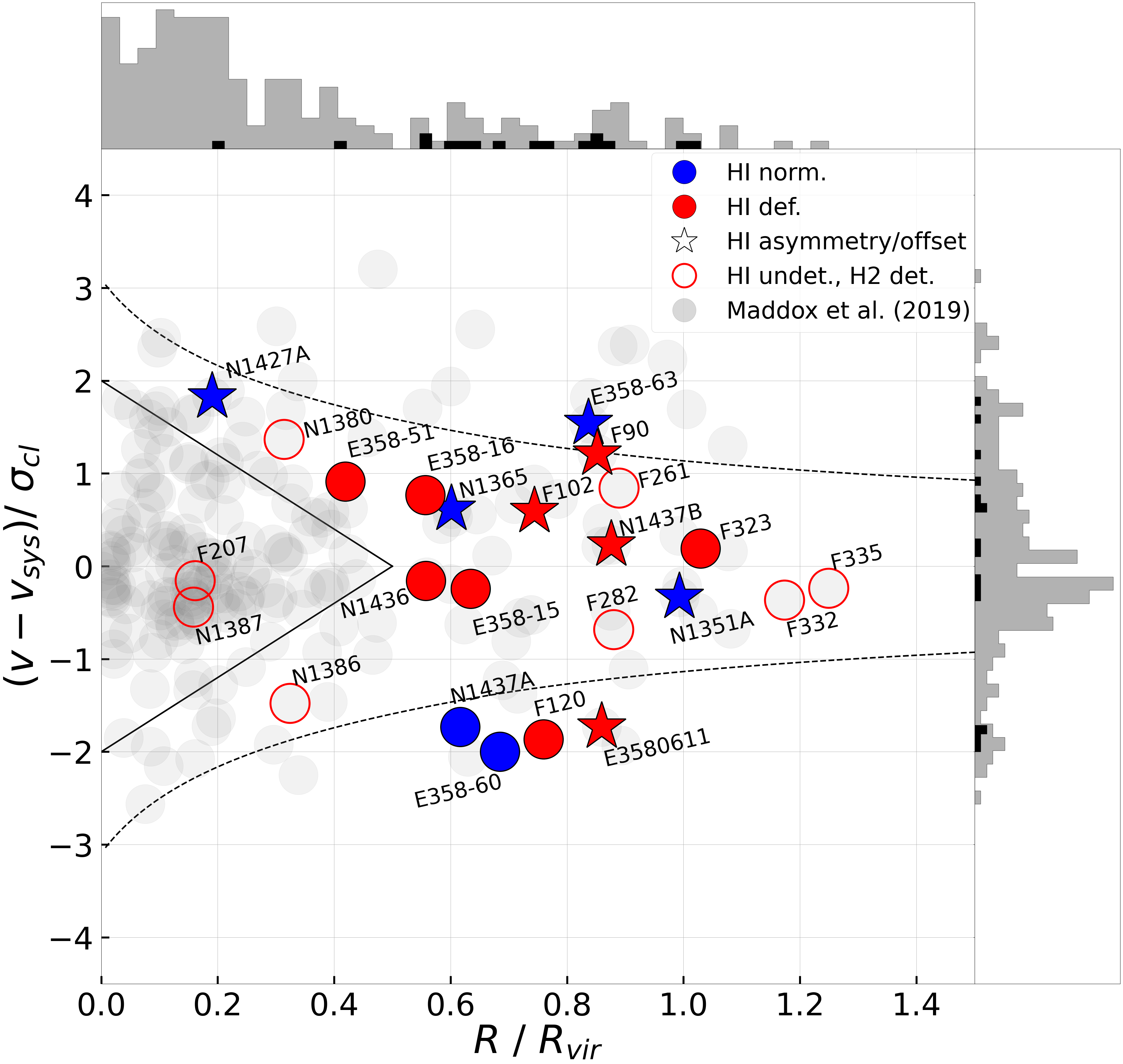}
        \caption{Phase space diagram of the Fornax cluster. Here, we used the same colour coding as Fig.~\ref{fig:distrib}. We show the caustic curves of the cluster with black dashed lines.
        Histograms: black and gray colours represent \hi{} detected galaxies and all Fornax members, respectively. The area within the black triangle shows the virialised area of the cluster}
        \label{fig:phsp}
    \end{figure}{}   

    Almost all our \hi{} detections (88$\%$ of the sample) are located farther than 0.5~$R_\mathrm{vir}$ in projection (the only exceptions being NGC~1427A at~0.2~$R_\mathrm{vir}$, and ESO~358-G051 at~0.4~$R_\mathrm{vir}$).
    In contrast, only 33$\%$ of all Fornax galaxies in our footprint are outside 0.5~$R_\mathrm{vir}$. 

    The same effect, with the addition of the kinematical information, can be seen in the projected phase-space diagram shown in Fig.~\ref{fig:phsp}. 
    We used a cluster systemic velocity of $v_\mathrm{sys}$~=~1442~\kms{} and a cluster velocity dispersion of $\sigma_\mathrm{cl}~=~318$~\kms{} (\citealp{Maddox2019}). 
    In order to draw caustic curves we proceeded as in \citet{jaffecaustic} using $M_\mathrm{vir}$~=~5$\times$10$^{13}$~M$_\odot$, $R_\mathrm{vir}$~=~700~kpc \citep{drinkwater2001substructure} and a cluster halo concentration parameter equal to 6 (\citealt{NFW}; the exact value of this parameter does not change the result significantly). The inner triangle shows the virialised area where it is more likely to find old members of the Fornax cluster \citep{rhee_2017ApJ}.
    From the right histogram in Fig.~\ref{fig:phsp}, we see that the spectroscopic Fornax members have a peak at the cluster systemic velocity, while velocities of our \hi{} detections cover all velocity range without any preferred velocity. Thus, both Fig.~\ref{fig:distrib} and Fig.~\ref{fig:phsp} show that the \hi{} galaxy sample does not follow the distribution of spectroscopic galaxies.
    
    In Fig.~\ref{fig:phsp}, we also see that 11 out of 16 galaxies are within the escape velocity boundary in projection (black lines). It is worth noting that the \hi{} deficient galaxies FCC~120 and ESO-LV~3580611 are located outside the caustic curves. 
    Among \hi{} disturbed galaxies, six out of eight are redshifted with respect to the recessional velocity of the cluster.
    
    Both Fig.~\ref{fig:distrib} and Fig.~\ref{fig:phsp} show that some galaxies in our sample are not just close by one another on the sky 
    but also in velocity. 
    The clearest case is that of NGC~1365 and its three neighbours: FCC~102, FCC~090 and ESO~358-G016. They are all within a region of 250~kpc and a velocity range of 200~\kms{}. 
    These galaxies might be part of a substructure which is accreting onto the cluster (as suggested by \citealt{drinkwater2001substructure}). In addition, most galaxies in this substructure exhibit indications of ongoing interactions with one another and/or with the intergalactic medium.
    Both \hi{} elongations in FCC~102 and NGC~1365 point in projection to the north, while \hi{} in FCC~090 is elongated towards the south. FCC~090, FCC~102 and NGC~1365 form a triplet of galaxies which lies along a line that points towards the centre of the cluster. ESO~358-G016 has a regular \hi{} distribution and it lies in projection to the north of NGC~1365. All but NGC~1365 are \hi{} deficient galaxies.
    The existence of a sub-group centred on NGC~1365 is also consistent with the large scale structure around Fornax (the Fornax-Eridanus supercluster), which is mainly made by several groups of galaxies that are assembling to form the cluster along the filament \citep{2011A&A...532A.104N}.
    
    On the east of the cluster centre, we also see that FCC~323 and NGC~1437B are near each other on the sky in projection ($\sim$120~kpc apart), as well as in velocity ($\sim$10~\kms{} difference). The unresolved \hi{} morphology of FCC~323 does not reveal, at this time, whether there is an ongoing interaction between those galaxies. We mark it as a potential subgroup.
    
    Another intriguing galaxy is ESO~358-G060. Among the galaxies with \mst{}$\leq10^9$~M$_\odot$ it is the only one with a normal \hi{} content compared to non-cluster galaxies (see Fig.~\ref{fig:himfraction}), and it shows no evidence of on-going interactions with the Fornax environment (see Fig.~\ref{fig:allmorph}). Given its position just outside the caustic curves in Fig.~\ref{fig:phsp}, we discuss in Sect.~\ref{sec:disc} whether it might not have entered the cluster yet.
    Another, similar case is NGC~1437A. It is a \hi{} rich galaxy which shows a quite regular \hi{} morphology. It is also just outside the caustic curves in Fig.~\ref{fig:phsp}. However, unlike the undisturbed ESO~358-G060, the optical appearance of NGC~1437A is peculiar, similar to that of NGC~1427A, since they both exhibit an arrow shaped optical morphology as pointed out in \cite{FDS_RAJ}. The MeerKAT Fornax Survey \citep{2016PaoloMeerkat} will deliver a higher resolution \hi{} image of this object and will be able to reveal any \hi{} asymmetries that might be hidden by projection effects within the ATCA beam.
    
 \section{Discussion}
 \label{sec:disc}
    The population of \hi{} detected Fornax galaxies exhibits several interesting features. Despite the limited resolution of our data, half of all detections reveal \hi{} asymmetries and offsets relative to the stellar body (Fig.~\ref{fig:allmorph}). Furthermore, the \hi{} sample as a whole is gas-poorer and is forming stars at a lower rate than samples of non$-$cluster galaxies in the same \mst{} range (Fig.~\ref{fig:himfraction} and Fig.~\ref{fig:sfr}), and half of the galaxies with \mst{}$>$10$^9$~M$_{\odot}$ have an anomalous \mhtwo{}/\mhi{} ratio. Finally, the \hi{} detections are distributed in a noticeably different way with respect to the majority of cluster galaxies both on the sky and in projected phase space (Fig.~\ref{fig:distrib} and Fig.~\ref{fig:phsp}). This body of evidence suggests that the Fornax environment is influencing the evolution of these galaxies, which may be the most recent arrivals in the cluster. In particular, the difference between the 3D distribution of \hi{} detected galaxies compared with that of spectroscopically confirmed Fornax members (\citealt{Maddox2019}), indicates that \hi{} is a crucial observable to test the volume of the cluster where \hi{} rich galaxies become \hi{} deficient, before a complete \hi{} removal in the inner part of the cluster.

    An outstanding question is how long it takes for a galaxy to lose its \hi{} -- the dominant component of the interstellar medium -- as it falls into a cluster. In the case of Fornax we can gain some insight through a joint analysis of Fig.~\ref{fig:deltasfr} and Fig.~\ref{fig:phsp}. If we assume that \hi{} is being actively removed from within galaxies in Fornax (as suggested by the frequently disturbed \hi{} morphologies - see Fig.~\ref{fig:allmorph}), the lack of \hi{} detections in the virialised region of Fig.~\ref{fig:phsp} implies that \hi{} removal happens on a time scale $\tau_\mathrm{HI,loss}$ shorter than the cluster's crossing time: $\tau_\mathrm{HI,loss} \leq \tau_\mathrm{cross}~\sim~R_\mathrm{vir}/\sigma_\mathrm{cl}~\sim~2$~Gyr (see Sect.~ \ref{subsec:distr}; this $\tau_\mathrm{cross}$ is similar to that estimated from simulations, 1.2$~\pm~$0.5~Gyr independent of cluster mass; \citealt{rhee_2017ApJ}.) Hence, we expect that our 16 detections will have lost most of their \hi{} by the time they reach the pericentre.
    
    On the other hand, Fig.~\ref{fig:deltasfr} suggests that, so far, \hi{} has been lost slowly in our \hi{} detections (having been removed and/or consumed, and not replenished). 
    In that figure, our Fornax \hi{} detections are distributed along the same correlation defined by non-cluster galaxies. Thus, for those Fornax galaxies, the SFR has so far had sufficient time to respond to a variation in \hi{} mass -- just like outside clusters. 
    Since the transition from \hi{} to new stars happens through the intermediate phase of \htwo{0pt}, this `equilibrium' between \hi{} and SFR implies that \htwo{0pt} is depleted faster than \hi{} is lost in galaxies currently at the cluster's outskirts, giving the entire cycle of \hi{}-to-\htwo{0pt}-to-SFR enough time to `see' the varying \hi{} content. Thus, while for Fornax as whole $\tau_\mathrm{HI,loss} \leq \tau_\mathrm{cross} \sim 2$~Gyr (see above), for our \hi{} detections in the outer regions of Fornax $\tau_\mathrm{HI,loss}$ has so far been $\geq~\tau_\mathrm{H_2,depl}~\sim~1\text{-}2$~Gyr \citep{bigiel_2008AJ}. (The \htwo{0pt} depletion time $\tau_\mathrm{H_2,depl}$ is only marginally shorter in Fornax, in particular at its outskirts, and anyway with large variations from galaxy to galaxy; \citealt{zabel2}.)
    
    It is likely that, further inside Fornax, \hi{} is removed faster than what we estimated above for galaxies at the cluster's outskirts. Indeed, simulations indicate that for a galaxy within $0.5~R_\mathrm{vir}$, $\tau_\mathrm{HI,loss}~<~0.5$~Gyr \citep{Antonino2016MNRAS.461.2630M}. Such a rapid \hi{} removal may not leave sufficient time for the SFR to `track' the decrease in \hi{} mass in Fig. \ref{fig:deltasfr}, resulting in galaxies moving to the left of the non-cluster sample. A confirmation of this effect may come from the eight \htwo{0pt}-detected galaxies (most of them \htwo{0pt}-deficient; \citealt{zabel}) where \hi{} has already been removed at least down to the ATCA \mhi{} sensitivity (left-pointing arrows in Fig.~\ref{fig:deltasfr}), and possibly by some \hi{} detections closer to the cluster centre (e.g. NGC 1436). 
    Indeed, these galaxies occupy a region to the left of the comparison sample, showing that their SFR is still significant despite their low \hi{} content. For these galaxies, \hi{} is likely to have been removed faster than \htwo{0pt} is depleted: $\tau_\mathrm{HI,loss} \leq \tau_\mathrm{H_2,depl}$. This conclusion is supported by the anomalously high \mhtwo{}/\mhi{} ratio of these galaxies in Fig.~\ref{fig:hionh2} (in the \mst{} range where we have a reliable comparison).
    
    Within the picture discussed above, \hi{}-detected galaxies in the outer regions of Fornax are thus first infallers, which are starting to interact with the Fornax environment and will lose most of their \hi{} by the time they reach the pericentre. Even at the current early stage of infall, they already show a relatively large diversity in \hi{} morphology (Fig.~\ref{fig:allmorph}) and mass (Fig.~\ref{fig:himfraction}). 
    More specifically, we found two morphologically undisturbed \hi{} rich galaxies ($12\%$ of the \hi{} detections);
    four morphologically disturbed \hi{} rich galaxies (25$\%$ of the \hi{} detections);
    four morphologically disturbed \hi{} deficient galaxies (25$\%$ of the \hi{} detections); six morphologically undisturbed - within the ATCA resolution - \hi{} deficient galaxies (38$\%$ of the \hi{} detections).
    
    The morphologically undisturbed \hi{}-rich galaxies are ESO~358-G060 and NGC~1437A. They both reside outside the caustic curves in projection (Fig.~\ref{fig:phsp}) and, given their low \mst{}, they should be easily perturbed by the Fornax environment. 
    Thus, we speculate that they are recent Fornax members which have not yet had enough time to be significantly affected by the cluster. One possible caveat is the low resolution of our images, which may hide \hi{} disturbances in particular in the case of the optically peculiar galaxy NGC~1437A. Another possibility is that they are outside the cluster volume or in a region with lower ICM density. 
    ESO~358-G060 is also an outlier in Fig.~\ref{fig:deltasfr}, where the relatively high \mhi{}/\mst{} does not correspond to a high SFR. In this galaxy no 
    SFR was detectable from WISE W3/W4 bands (after subtracting the old stellar population light), and the NUV contribution to star formation is low. 
    Several physical processes might account for the low SFR with respect to the large \hi{} reservoir: for example, an inefficient \hi{} to \htwo{0pt} conversion due to a small amount of dust, a large angular momentum which prevents the \hi{} from collapsing, and/or an \hi{} external origin (e.g. see \citealt{2016MNRAS.462..382G,2018MNRAS.476..896G}).
    
    The morphologically disturbed \hi{}-rich galaxies are ESO~358-G063, NGC~1351A, NGC~1427A, NGC~1365. We discuss the last galaxy when we focus our attention on the \hi{} detected subgroup of interacting galaxies. 
    ESO~358-G063 and NGC~1351A, are both disc galaxies north of NGC~1399. They are both quite isolated from all other Fornax Galaxies in RA, Dec and velocity in projection (see Fig.~\ref{fig:distrib} and Fig.~\ref{fig:phsp}). Thus, their slightly \hi{} disturbed morphologies described in Sect.\ref{sec:emiss} might be due to the interaction with the ICM of the Fornax cluster. 
    Furthermore, the agreement of \mhtwo{}/\mhi{} of ESO~358-G063 and NGC~1351A  with the xGASS scaling relation in Fig.~\ref{fig:hionh2} suggests that they are recent Fornax members, thus the cluster environment has not had enough time to significantly deplete their \hi{} reservoirs.   
    The last morphologically disturbed \hi{} rich galaxy is NGC~1427A. This is the galaxy with the second-highest \hi{} mass in our sample of Fornax \hi{} detections.
    In Fig.~\ref{fig:distrib} and Fig.~\ref{fig:phsp}, we see that NGC~1427A is the closest \hi{} detected galaxy to the centre of the cluster in projection, but it has also the highest velocity. Thus, it may be a new Fornax member which is infalling from the foreground. 
    As already mentioned, \cite{karenNGC1427A} studied the origin of the \hi{} tail using the same data we present here.
    They concluded that, unlike previously suggestions, ram-pressure is unlikely to be the main process shaping the galaxy's optical appearance. Instead, NGC~1427A is most likely a recent merger remnant, thus shaped by tidal forces.
    The recent merger might be the cause of the low molecular column density and/or its low metallicity, resulting currently undetected by ALMA (\citealt{zabel}; it is the upper limit with the highest \mst{} in Fig.~\ref{fig:hionh2}).
    NGC~1427A is also the second \hi{} rich outlier with low SFR in Fig.~\ref{fig:deltasfr}. The \hi{}-to-\htwo{0pt} conversion might be inefficient to have a SFR consistent with its \mhi{}/\mst{} ratio.
    
    The only \hi{} deficient and morphologically disturbed galaxies which do not belong to the NGC~1365 subgroup are: ESO-LV~3580611 and NGC~1437B. They are very close to one another on the sky but have significantly different velocities.
    The former is outside the caustic curve in projection (Fig.~\ref{fig:phsp}), which supports the hypothesis of an new infalling Fornax member made by \cite{Schroder2001}.
    The latter has a velocity similar to the recessional velocity of the cluster.
    \hi{} and molecular morphologies (the latter detected by \citealt{zabel}) are elongated in the same direction. This suggests that both gas phases are experiencing the same environmental interaction. \citealp{FDS_RAJ} detected a tidal tail in NGC~1437B, which  may be due to a recent fly-by of another galaxy. As mentioned in Sect.\ref{subsec:distr}, although the evidence is not strong, NGC~1437B might be part of a subgroup of interacting galaxies which includes FCC~323. Thus, FCC~323 might be the fly-by galaxy that NGC~1437B has interacted with.

    It is difficult to comment on the morphologically undisturbed \hi{} deficient galaxies, since their symmetric \hi{} distribution may be a consequence of the ATCA resolution. However, a very peculiar case is the truncated \hi{} disc of NGC~1436.
    It is the closest spiral galaxy to the centre of the cluster detected in \hi{}. \cite{FDS_RAJ} observed an ongoing morphological transition into lenticular: the spiral structure is found only in its inner region, while the outer disc has the smooth appearance typically found in S0 galaxies.

    The inner part of NGC~1436 appears regular also in \htwo{0pt}, and the galaxy is just moderately \htwo{0pt} deficient \citep{zabel}.
    It is also the only galaxy detected both in \hi{} and \htwo{0pt} which shows a high \mhtwo{}/\mhi{} ratio in Fig.~\ref{fig:hionh2}. 
    These results and the evidence of morphological distortions only in the outer part of the galaxy suggest that this galaxy may have gone through a quick interaction with the cluster environment, which did not affect the inner spiral structure yet. This idea is corroborated by the fact that NGC~1436 lies on the upper edge on the comparison sample in Fig.~\ref{fig:deltasfr}, which means that despite its \hi{} deficiency it is still forming stars at a significant rate.
    
    Finally, we focus on the NGC~1365 subgroup. We discussed all the \hi{} morphologies and the 3D distribution of the subgroup members in Sect.\ref{sec:hiproducts} and Sect.\ref{subsec:distr}, respectively. As mentioned earlier, NGC~1365 is the only \hi{} rich galaxy of the subgroup (Fig.~\ref{fig:himfraction}) with a \mhi{} at least two orders of magnitude larger than the other members. The \hi{} distribution both in NGC~1365 and in FCC~102 is elongated to the north, while it is elongated to the south in FCC~090. ESO~358-G016 is the only galaxy with a regular \hi{} morphology. It is also the only galaxy located north of NGC~1365. We propose two scenarios in order to explain the properties of the subgroup and its members:
    the former is a case of interaction between galaxies and the cluster environment which is responsible of the high \mhi{}/\mst{} ratio of the low mass members. In contrast, due to to its deep gravitational potential, NGC~1365 has been able to retain its \hi{}, although some of it has been perturbed.
    The latter scenario we propose, it is a case of preprocessing in a group of galaxies where the local environment of the subgroup was able to affect the \mhi{}/\mst{} ratio of the low mass members before the group began to interact with the cluster environment.
    
    In general, although we found a large variety of \hi{} properties in our sample of \hi{} detections, we detected an overall trend towards \hi{} disturbances and deficiency in Fornax.
    Fig.~\ref{fig:himfraction} makes evident an already evolved state of Fornax \hi{} galaxies where $\sim$2/3 of the galaxies are \hi{} deficient. Fig.~\ref{fig:himfraction} also shows that the Fornax environment is more effective in altering the gas content of galaxies with \mst{}~$<$~3~$\times$~10$^9$~M$_{\odot}$) (see Sect.\ref{subse:himfrac}).

    \cite{zabel} presented a similar study, where they compare Fornax and field galaxies based on their molecular gas properties. They found some molecular deficiency in all their detections except NGC~1365. However, galaxies with \mst{}~$<$~3$\times$10$^9$~M$_{\odot}$ are both more \htwo{0pt}-deficient and morphologically disturbed with respect to more massive galaxies, whose molecular gas morphology is always regular. 
    We do not observe such a clear difference in \hi{}.
    Indeed, we observe disturbed \hi{} morphologies across our entire \mst{} range, confirming that atomic hydrogen is the best tracer of early interactions.
    Despite this difference between \hi{} and \htwo{0pt} morphologies, we also note some similarities between our results and those in \cite{zabel}. Indeed, the mass range of molecular disturbed galaxies is also characterised by a stronger \hi{} depletion.
    Conversely, the \hi{} depletion is weaker in the mass range in which galaxies show regular molecular-gas morphologies (\mst{}~$>$~3$\times$10$^9$~M$_{\odot}$; Sect.\ref{subse:himfrac}). 
    Thus, both gas phases show that the Fornax environment is more effective in altering the gas content of low-mass galaxies compared to high-mass galaxies.
    
    Finally, our results are in agreement with the FDS$\&$F3D results \citep[]{iodiceFDS2019, 2019A&A...623A...1I}. Indeed, almost all galaxies with a disturbed \hi{} morphology are of late type and belong to the group of the infalling galaxies in \citep{iodiceFDS2019}, which are symmetrically distributed around the cluster's central region. These galaxies have active star formation and are located in the low-density region of the cluster, where the X-ray emission is faint or absent. Our results show that they are interacting with the cluster environment.
    Deeper into the cluster, the lack of \hi{} detections is consistent with the result that this region is dominated by evolved early-type galaxies.
    Some of these galaxies have been able to retain part of their \htwo{0pt} reservoirs \citep{zabel}, but not their \hi{}.

 \section{Summary}
\label{sec:summar}
    The blind ATCA \hi{} survey of the Fornax galaxy cluster covers a field of 15~deg$^2$ out to a distance of $\sim R_\mathrm{vir}$ from the cluster centre. It has a spatial and velocity resolution of 67\arcsec~$\times$~95\arcsec and 6.6~\kms{}, respectively, and a 3$\sigma$ \nhi{} and \mhi{} sensitivity of ~$\sim$2$~\times$~10$^{19}$cm$^{-2}$ and $\sim$2$~\times~$~10$^7$M$_\odot$, respectively. The survey revealed \hi{} emission from 16 Fornax galaxies covering a mass range of about three orders of magnitude, from $8\times 10^6$ to $1.5\times 10^{10}$~M$_\odot$. These galaxies exhibit a variety of disturbances of the \hi{} morphology, including asymmetries, tails, offsets between \hi{} and optical centres and a case of a truncated \hi{} disc (Fig.~\ref{fig:allmorph}). This suggests environmental interactions within or on their way to Fornax (whether with other galaxies, the large-scale potential or the intergalactic medium), supported by the offset of Fornax galaxies towards low \mhi{}/\mst{} ratios with respect to the xGASS \mst{}-\mhi{}/\mst{} scaling relation (Fig.~\ref{fig:himfraction}), and resulting in \hi{} deficiencies similar to those observed in the Virgo cluster (Fig.~\ref{fig:hivirgo}). 
    The \hi{} sample of Fornax galaxies is also forming stars at a lower rate than samples of non-cluster galaxies at fixed \mst{} (Fig.~\ref{fig:sfr}). This deficit of SFR is consistent with the deficit of \hi{} when compared to non-cluster galaxies (Fig.~\ref{fig:deltasfr}).
    
    Our 16 detections reside outside the virialised region of the cluster -- where the distribution of the general population of Fornax galaxies is clustered -- both on the sky and in the projected phase space diagram (Fig.~\ref{fig:distrib} and Fig.~\ref{fig:phsp}). This result implies that \hi{} is lost down to the ATCA sensitivity within a crossing time ($\tau_\mathrm{HI,loss}~\leq~\tau_\mathrm{cross}~\sim2$~Gyr), and that  our \hi{} detections are recent arrivals in the cluster. They still reside at the outskirts of Fornax, where their \hi{} and SFR properties suggest that \hi{} has so far been lost on a time scale longer than the \htwo{0pt} depletion time ($\tau_\mathrm{HI,loss}~\geq~\tau_\mathrm{H_2,depl}~\sim1\mathrm{-}2$~Gyr). In the cluster's central regions \hi{} removal is likely to proceed faster ($\tau_\mathrm{HI,loss}~<~\tau_\mathrm{H_2,depl}$). This is supported by the relatively high SFR of \hi{}-undetected, \htwo{0pt}-detected galaxies and by the anomalously high \mhtwo{}/\mhi{} ratios of galaxies in those regions (Fig.~\ref{fig:hionh2}, Fig.~\ref{fig:distrib} and Fig.~\ref{fig:phsp}). These are galaxies where SFR is likely to be proceeding relatively unperturbed after rapid removal of the \hi{}.
    
    This picture is enriched by the new detection of the NGC~1365 subgroup -- where both pre-processing and early interaction with the cluster environment are plausible scenarios to account for the \hi{} properties of its members -- and by the detection of several galaxies with peculiar ISM properties, such as some \hi{}-rich but \htwo{0pt}-poor and low-SFR galaxies (NGC~1427A, ESO~358-G060). The future MeerKAT Fornax Survey \citep{2016PaoloMeerkat} will observe this cluster with a better resolution and sensitivity than those of our ATCA survey, enabling a further step forward in the study of the evolution of Fornax galaxies.

\begin{acknowledgements}

This project has received funding from the European Research Council (ERC) under the European Union’s Horizon 2020 research and innovation programme (grant agreement no. 679627; project name FORNAX). Parts of this research were supported by the Australian Research Council Centre of Excellence for All Sky Astrophysics in 3 Dimensions (ASTRO 3D), through project number CE170100013. LC is the recipient of an Australian Research Council Future Fellowship (FT180100066) funded by the Australian Government. This publication has received funding from the European Union Horizon 2020 research and innovation programme under the Marie Skłodowska-Curie grant agreement number 721463 to the SUNDIAL ITN network. NZ acknowledges support from the European Research Council (ERC) in the form of Consolidator Grant CosmicDust (ERC-2014-CoG-647939). This work made use of the Digitized Sky Surveys, which were produced at the Space Telescope Science Institute under U.S. Government grant NAG W-2166. The Australia Telescope Compact Array is part of the Australia Telescope National Facility which is funded by the Australian Government for operation as a National Facility managed by CSIRO. We acknowledge the Gomeroi people as the traditional owners of the Observatory site.

\end{acknowledgements}

\newpage
\onecolumn

\begin{sidewaystable}[p]
    \centering 
    \small\addtolength{\tabcolsep}{-1pt}
    \footnotesize{                           
    \begin{tabular}{llrrrrrrrrrrrc}             
    \toprule                                 
         Name   &  FCC    &      RA           &         Dec         & \vspe{}       & \vbary{}     &     Flux         &   \mhi{}                 &  M$_\star$                         &     RMS              &           Flux     &  Morph.    & D$_{25}$(B) &     Notes                           \\ 
                                                                                                                                                                                                                                                                             
                &         &   \scriptsize{(J2000)}   &     \scriptsize{(J2000)}        &                &               &                  &                           &                                    &                      &           \scriptsize{(literature)}   &                &                     &                   \\

                &         &   (hh:mm:ss.ss)         &         (dd:mm:ss.ss)     & $\left(\frac{\mathrm{km}}{\mathrm{s}}\right)$       & $\left(\frac{\mathrm{km}}{\mathrm{s}}\right)$      &  $\left(\frac{\mathrm{Jy km}}{\mathrm{s}}\right)$       &   (10$^8$M$\odot$)    &  (10$^8$M$_\odot$)             & ($\frac{\mathrm{mJy}}{\mathrm{beam}}$)              &        $\left(\frac{\mathrm{Jy km}}{\mathrm{s}}\right)$    &     $ $           &  (arcsec)   &        $ $    \\

    \midrule                                                                                          
    
ESO 358-G015   &  113 &  03:33:06.85 &  -34:48:29.19 &  1365 &  1389 &   1.0$\pm$0.2 & 0.9$\pm$0.2     &    8$\pm$2        &  2.4 & 1.4$\pm$0.2$^\ast$  & Scd &   72.10    & \hi{} def.   \\
ESO 358-G016   &  115 &  03:33:09.19 &  -35:43:06.69 &  1686 &  1694 & 0.8$\pm$0.2   & 0.7$\pm$0.2     &   1.2$\pm$0.2     &  2.5 & 1.2$\pm$0.5$^\triangle$ &  Im &   48.42   &  \hi{} def.;   N1365 s.group   \\
ESO 358-G051   &  263 &  03:41:32.59 &  -34:53:17.99 &  1733 &  1731 &   1.1$\pm$0.2 & 1.0$\pm$0.2     &    21$\pm$5       &  2.5 & 2.8$\pm$0.4$^\circ$ & Scd &   92.90  & \hi{} def.   \\
ESO 358-G060   &  302 &  03:45:12.14 &  -35:34:15.26 &   806 &   803 &  12$\pm$2     &  11$\pm$2       &  1.0$\pm$0.6      &  2.4 &   10.6$\pm$0.6$^\blacktriangledown$  &   Scd &  104.30   &    \\
ESO 358-G063   &  312 &  03:46:19.00 &  -34:56:36.80 &  1932 &  1940 &  18$\pm$4     &  17$\pm$3       &  110$\pm$30       &  2.6 &      18.7$\pm$0.9$^\blacktriangle$  & Scd &  280.60   & \hi{} dist;    \\
ESO-LV 3580611 &  306 &  03:45:45.39 &  -36:20:47.50 &   891 &   894 &   1.1$\pm$0.2 & 1.1$\pm$0.2     &  1.0$\pm$0.4      &  2.4 & 2.7$\pm$0.7$^\triangledown$ & Im &   31.99   & \hi{} def.; \hi{} dist;    \\
FCC 090        &  090 &  03:31:08.26 &  -36:17:24.50 &  1827 &  1823 & 0.6$\pm$0.2   & 0.6$\pm$0.1     &   13 $\pm$3       &  2.3 & \# &   E &   62.80 & \hi{} def.; \hi{} dist;   N1365 s.group   \\
FCC 102        &  102 &  03:32:10.73 &  -36:13:14.91 & 1722  & 1631  & 0.5$\pm$0.1   & 0.5$\pm$0.1     &   1.6$\pm$0.5     &  2.5 & \# & Im &  33.50  & \hi{} def.,\hi{} dist;   N1365 s.group   \\
FCC 120        &  120 &  03:33:34.22 &  -36:36:21.29 &  849  &  846  & 0.5$\pm$0.1   & 0.4$\pm$0.1     &   1.6$\pm$0.5     &  2.4 & 1.9$\pm$0.9$^\triangledown$ & Im &   48.13 & \hi{} def.   \\
FCC 207        &  207 &  03:38:19.27 &  -35:07:44.69 &  1393 & \#    &  $<$0.1       &  $<$0.1         &  5  $\pm$ 1       &  2.3 & \# & E   &   36.65   & \hi{} undet.  \\
FCC 261        &  261 &  03:41:21.52 &  -33:46:09.19 &  1710 & \#    &  $<$0.1       &  $<$0.1         &  3.6  $\pm$ 0.9   &  2.3 & \# & E   &   39.75   & \hi{} undet.  \\
FCC 282        &  282 &  03:42:45.31 &  -33:55:13.80 &  1225 & \#    &  $<$0.2       &  $<$0.2         &  18 $\pm$ 4       &  2.5 & \# & E   &   54.39   & \hi{} undet.  \\
FCC 323        &  323 &  03:47:37.52 &  -36:21:46.83 &  \#   &  1502 &  0.08$\pm$0.03 &  0.08$\pm$0.03 &  0.6  $\pm$ 0.6   &  2.4 & \# & E   &   \#   & \hi{} def.   \\
FCC 332        &  332 &  03:49:49.02 &  -35:56:44.09 &  1326 & \#    &  $<$0.1       &  $<$0.1         &  7  $\pm$ 2       &  2.3 & \# & E   &   43.90   & \hi{} undet.  \\
FCC 335        &  335 &  03:50:36.73 &  -35:54:33.59 &  1367 & \#    &  $<$0.1       &  $<$0.1         &  27 $\pm$ 6       &  2.6 & \# & E   &   80.90   & \hi{} undet.  \\
NGC 1351A      &  067 &  03:28:48.72 &  -35:10:41.30 &  1336 &  1337 &   5$\pm$1     &  5$\pm$1        &     35$\pm$8      &  3.2 & 6.1$\pm$0.9$^\blacktriangle$ &  Sc &  161.50   & \hi{} dist;    \\
NGC 1365       &  121 &  03:33:36.37 &  -36:08:25.44 &  1638 &  1641 &  160$\pm$30   &  150$\pm$30     & 600$\pm$100       &  2.8 &  146$\pm$3$^\blacktriangledown$ &  Sb &  673.20   &  \hi{ dist.;}   N1365 s.group\\
NGC 1380       &  167 &  03:36:27.59 &  -34:58:34.41 &  1877 & \#    &  $<$0.6       &  $<$0.5         &  1400 $\pm$ 300   &  2.6 & \# & S0a &   287.20   & \hi{} undet.  \\
NGC 1386        &  179 &  03:36:46.18 &  -35:59:57.86 &  972  & \#    &  $<$0.5       &  $<$0.5         &  120  $\pm$ 30    &  2.9 & \# & Sa  &   203.30   & \hi{} undet.  \\
NGC 1387       &  184 &  03:36:57.06 &  -35:30:23.90 &  1302 & \#    &  $<$0.3       &  $<$0.3         &  900  $\pm$ 200   &  2.1 & \# & E   &   169.10   & \hi{} undet.  \\  
NGC 1427A      &  235 &  03:40:09.30 &  -35:37:27.99 &  2023 &  2026 &  22$\pm$4     &  21$\pm$4       & 23$\pm$5          &  2.3 &   22.5$\pm$0.8$^\blacktriangledown$  &  Im &  140.70    & \hi{} dist;     \\
NGC 1436       &  290 &  03:43:37.08 &  -35:51:10.90 & 1392  & 1414  & 0.6$\pm$0.2   & 0.6$\pm$0.2     &  160$\pm$40       &  2.5 & 1.7$\pm$0.2$^\diamond$ &  Sc &  177.10   & \hi{} def.; truncated \hi{} disc   \\
NGC 1437A      &  285 &  03:43:02.19 &  -36:16:24.14 &   891 &   887 &   6$\pm$1     &   6$\pm$1       &     10$\pm$3      &  2.4 & 7$\pm$1$^\blacktriangle$   &   Scd &  111.70  &    \\
NGC 1437B      &  308 &  03:45:54.85 &  -36:21:25.09 &  1515 &  1501 &   2.6$\pm$0.5 & 2.4$\pm$0.5     &  50  $\pm$10      &  2.5 & 3.2$\pm$0.6$^\blacktriangle$ & Scd &  157.80   & \hi{} def.; \hi{} dist; \\
\bottomrule
\end{tabular}
}
\caption{We report parameters on the \hi{} detected galaxies. RA and Dec coordinates correspond to the optical centres from NED. Errors on the total \hi{} fluxes are evaluated as described in Sect.\ref{sec:emiss}. `Flux (literature)' shows the total fluxes measured from other surveys as follows: $\diamond$ \citet{courtoisGBT}; $\triangle$ \citet{Bureau1996}; $\ast$ \citet{ESO015_1998AJ....116.1169M}; $\circ$ \citet{ESO051_1998A&AS..130..333T}; $\triangledown$ \citet{Schroder2001}; $\blacktriangledown$ \citet{BaerbelBGC2004}; $\blacktriangle$ HIPASS data reprocessed by us. The total flux uncertainty of the comparison spectra are calculated as described in Sect.~\ref{sec:emiss}  combining the noise in the spectrum and the flux-scale uncertainty of each survey except for ESO~358-G016 for which the flux-scale uncertainty was not provided. For this galaxy we show the uncertainty on the total flux given in \citet{Bureau1996}. Columns `Morph.' and `D$_{25}$(B)' show the optical morphology and the optical size of our galaxies used in Sect.~\ref{subse:himfrac} to evaluate the expected \hi{} content of our galaxies with the \citeauthor{H&G84}'s (\citeyear{H&G84}) method. D$_{25}$(B) is the optical isophotal diameter measured at 25 mag/arcsec$^2$ in B-band; we collected these values from NED for 17 out of 24 galaxies. Almost all of the optical diameters comes from RC3 catalogue \citep{1991rc3..book.....D}, two out of 15 are from \cite{1989spce.book.....L}. We estimated D$_{25}$(B) for all the remaining galaxies except for FCC~323 as follows: from the optical multi-component decomposition made by \cite{2021arXiv210105699S}, we scaled the r-band radial profile to B-band. In order to do this, we previously converted g-r from \cite{venhola2019A&A} to obtain B-r using the intermediate conversion formula g-r to B-g \citep{lupton2005}. In column `Notes': \hi{}~def, \hi{}~dist, N~1365~s.group stand for \hi{}~deficient, \hi{} morphologically disturbed galaxy, and galaxy included in the NGC~1365 subgroup, respectively.}\label{tab:mass}

\end{sidewaystable}

\twocolumn
\bibliographystyle{aa} 
\bibliography{biblio2.bib} 

\begin{thebibliography}{102}
\expandafter\ifx\csname natexlab\endcsname\relax\def\natexlab#1{#1}\fi

\bibitem[{{Barnes} {et~al.}(2001){Barnes}, {Staveley-Smith}, {de Blok},
  {Oosterloo}, {Stewart}, {Wright}, {Banks}, {Bhathal}, {Boyce}, {Calabretta},
  {Disney}, {Drinkwater}, {Ekers}, {Freeman}, {Gibson}, {Green}, {Haynes}, {te
  Lintel Hekkert}, {Henning}, {Jerjen}, {Juraszek}, {Kesteven}, {Kilborn},
  {Knezek}, {Koribalski}, {Kraan-Korteweg}, {Malin}, {Marquarding}, {Minchin},
  {Mould}, {Price}, {Putman}, {Ryder}, {Sadler}, {Schr{\"o}der}, {Stootman},
  {Webster}, {Wilson}, \& {Ye}}]{HIPASS_2001MNRAS}
{Barnes}, D.~G., {Staveley-Smith}, L., {de Blok}, W.~J.~G., {et~al.} 2001,
  \mnras, 322, 486

\bibitem[{{Barnes} {et~al.}(1997){Barnes}, {Staveley-Smith}, {Webster}, \&
  {Walsh}}]{Barnes1997fornaxcluster}
{Barnes}, D.~G., {Staveley-Smith}, L., {Webster}, R.~L., \& {Walsh}, W. 1997,
  \mnras, 288, 307

\bibitem[{{Bigiel} {et~al.}(2008){Bigiel}, {Leroy}, {Walter}, {Brinks}, {de
  Blok}, {Madore}, \& {Thornley}}]{bigiel_2008AJ}
{Bigiel}, F., {Leroy}, A., {Walter}, F., {et~al.} 2008, \aj, 136, 2846

\bibitem[{Blakeslee {et~al.}(2009)Blakeslee, Jord{\'{a}}n, Mei,
  C{\^{o}}t{\'{e}}, Ferrarese, Infante, Peng, Tonry, \&
  West}]{blakeslee2009acs}
Blakeslee, J.~P., Jord{\'{a}}n, A., Mei, S., {et~al.} 2009, The Astrophysical
  Journal, 694, 556

\bibitem[{{Boquien} {et~al.}(2016){Boquien}, {Kennicutt}, {Calzetti}, {Dale},
  {Galametz}, {Sauvage}, {Croxall}, {Draine}, {Kirkpatrick}, {Kumari}, {Hunt},
  {De Looze}, {Pellegrini}, {Rela{\~n}o}, {Smith}, \&
  {Tabatabaei}}]{boquien2016A}
{Boquien}, M., {Kennicutt}, R., {Calzetti}, D., {et~al.} 2016, \aap, 591, A6

\bibitem[{{Boselli} {et~al.}(2014){Boselli}, {Cortese}, \&
  {Boquien}}]{HRSboselli2014}
{Boselli}, A., {Cortese}, L., \& {Boquien}, M. 2014, \aap, 564, A65

\bibitem[{{Boselli} \& {Gavazzi}(2006)}]{boselligavazzi}
{Boselli}, A. \& {Gavazzi}, G. 2006, \pasp, 118, 517

\bibitem[{{Boselli} \& {Gavazzi}(2009)}]{BoselliGavazzi2009}
{Boselli}, A. \& {Gavazzi}, G. 2009, \aap, 508, 201

\bibitem[{{Bureau} {et~al.}(1996){Bureau}, {Mould}, \&
  {Staveley-Smith}}]{Bureau1996}
{Bureau}, M., {Mould}, J.~R., \& {Staveley-Smith}, L. 1996, \apj, 463, 60

\bibitem[{{Cantiello} {et~al.}(2018){Cantiello}, {D'Abrusco}, {Spavone},
  {Paolillo}, {Capaccioli}, {Limatola}, {Grado}, {Iodice}, {Raimondo},
  {Napolitano}, {Blakeslee}, {Brocato}, {Forbes}, {Hilker}, {Mieske},
  {Peletier}, {van de Ven}, \& {Schipani}}]{2018A&A...611A..93C}
{Cantiello}, M., {D'Abrusco}, R., {Spavone}, M., {et~al.} 2018, \aap, 611, A93

\bibitem[{{Catinella} {et~al.}(2018){Catinella}, {Saintonge}, {Janowiecki},
  {Cortese}, {Dav{\'e}}, {Lemonias}, {Cooper}, {Schiminovich}, {Hummels},
  {Fabello}, {Ger{\'e}b}, {Kilborn}, \& {Wang}}]{BarbaraxGASS}
{Catinella}, B., {Saintonge}, A., {Janowiecki}, S., {et~al.} 2018, \mnras, 476,
  875

\bibitem[{{Cayatte} {et~al.}(1994){Cayatte}, {Kotanyi}, {Balkowski}, \& {van
  Gorkom}}]{1994AJ....107.1003C}
{Cayatte}, V., {Kotanyi}, C., {Balkowski}, C., \& {van Gorkom}, J.~H. 1994,
  \aj, 107, 1003

\bibitem[{{Chamaraux} {et~al.}(1980){Chamaraux}, {Balkowski}, \&
  {Gerard}}]{1980A&A....83...38C}
{Chamaraux}, P., {Balkowski}, C., \& {Gerard}, E. 1980, \aap, 83, 38

\bibitem[{{Chung} {et~al.}(2009){Chung}, {van Gorkom}, {Kenney}, {Crowl}, \&
  {Vollmer}}]{ChungVIVA2009}
{Chung}, A., {van Gorkom}, J.~H., {Kenney}, J. D.~P., {Crowl}, H., \&
  {Vollmer}, B. 2009, \aj, 138, 1741

\bibitem[{{Cluver} {et~al.}(2017){Cluver}, {Jarrett}, {Dale}, {Smith},
  {August}, \& {Brown}}]{cluver2017ApJ}
{Cluver}, M.~E., {Jarrett}, T.~H., {Dale}, D.~A., {et~al.} 2017, \apj, 850, 68

\bibitem[{{Cluver} {et~al.}(2014{\natexlab{a}}){Cluver}, {Jarrett}, {Hopkins},
  {Driver}, {Liske}, {Gunawardhana}, {Taylor}, {Robotham}, {Alpaslan},
  {Baldry}, {Brown}, {Peacock}, {Popescu}, {Tuffs}, {Bauer}, {Bland -Hawthorn},
  {Colless}, {Holwerda}, {Lara-L{\'o}pez}, {Leschinski},
  {L{\'o}pez-S{\'a}nchez}, {Norberg}, {Owers}, {Wang}, \& {Wilkins}}]{2014wise}
{Cluver}, M.~E., {Jarrett}, T.~H., {Hopkins}, A.~M., {et~al.}
  2014{\natexlab{a}}, \apj, 782, 90

\bibitem[{{Cluver} {et~al.}(2014{\natexlab{b}}){Cluver}, {Jarrett}, {Hopkins},
  {Driver}, {Liske}, {Gunawardhana}, {Taylor}, {Robotham}, {Alpaslan},
  {Baldry}, {Brown}, {Peacock}, {Popescu}, {Tuffs}, {Bauer}, {Bland -Hawthorn},
  {Colless}, {Holwerda}, {Lara-L{\'o}pez}, {Leschinski},
  {L{\'o}pez-S{\'a}nchez}, {Norberg}, {Owers}, {Wang}, \&
  {Wilkins}}]{cluver2014ApJ}
{Cluver}, M.~E., {Jarrett}, T.~H., {Hopkins}, A.~M., {et~al.}
  2014{\natexlab{b}}, \apj, 782, 90

\bibitem[{{Condon} {et~al.}(1998){Condon}, {Cotton}, {Greisen}, {Yin},
  {Perley}, {Taylor}, \& {Broderick}}]{NVSS_1998AJ}
{Condon}, J.~J., {Cotton}, W.~D., {Greisen}, E.~W., {et~al.} 1998, \aj, 115,
  1693

\bibitem[{{Cortese} {et~al.}(2016){Cortese}, {Bekki}, {Boselli}, {Catinella},
  {Ciesla}, {Hughes}, {Baes}, {Bendo}, {Boquien}, {de Looze}, {Smith},
  {Spinoglio}, \& {Viaene}}]{2016MNRAS.459.3574C}
{Cortese}, L., {Bekki}, K., {Boselli}, A., {et~al.} 2016, \mnras, 459, 3574

\bibitem[{{Cortese} {et~al.}(2011){Cortese}, {Catinella}, {Boissier},
  {Boselli}, \& {Heinis}}]{LucaVirgo2011}
{Cortese}, L., {Catinella}, B., {Boissier}, S., {Boselli}, A., \& {Heinis}, S.
  2011, \mnras, 415, 1797

\bibitem[{{Courtois} {et~al.}(2009){Courtois}, {Tully}, {Fisher}, {Bonhomme},
  {Zavodny}, \& {Barnes}}]{courtoisGBT}
{Courtois}, H.~M., {Tully}, R.~B., {Fisher}, J.~R., {et~al.} 2009, \aj, 138,
  1938

\bibitem[{{Davies} \& {Lewis}(1973)}]{1973MNRAS.165..231D}
{Davies}, R.~D. \& {Lewis}, B.~M. 1973, \mnras, 165, 231

\bibitem[{{de Vaucouleurs} {et~al.}(1991){de Vaucouleurs}, {de Vaucouleurs},
  {Corwin}, {Buta}, {Paturel}, \& {Fouque}}]{1991rc3..book.....D}
{de Vaucouleurs}, G., {de Vaucouleurs}, A., {Corwin}, Herold~G., J., {et~al.}
  1991, {Third Reference Catalogue of Bright Galaxies}

\bibitem[{{Dey} {et~al.}(2019){Dey}, {Schlegel}, {Lang}, {Blum}, {Burleigh},
  {Fan}, {Findlay}, {Finkbeiner}, {Herrera}, {Juneau}, {Landriau}, {Levi},
  {McGreer}, {Meisner}, {Myers}, {Moustakas}, {Nugent}, {Patej}, {Schlafly},
  {Walker}, {Valdes}, {Weaver}, {Y{\`e}che}, {Zou}, {Zhou}, {Abareshi},
  {Abbott}, {Abolfathi}, {Aguilera}, {Alam}, {Allen}, {Alvarez}, {Annis},
  {Ansarinejad}, {Aubert}, {Beechert}, {Bell}, {BenZvi}, {Beutler}, {Bielby},
  {Bolton}, {Brice{\~n}o}, {Buckley-Geer}, {Butler}, {Calamida}, {Carlberg},
  {Carter}, {Casas}, {Castander}, {Choi}, {Comparat}, {Cukanovaite}, {Delubac},
  {DeVries}, {Dey}, {Dhungana}, {Dickinson}, {Ding}, {Donaldson}, {Duan},
  {Duckworth}, {Eftekharzadeh}, {Eisenstein}, {Etourneau}, {Fagrelius},
  {Farihi}, {Fitzpatrick}, {Font-Ribera}, {Fulmer}, {G{\"a}nsicke},
  {Gaztanaga}, {George}, {Gerdes}, {Gontcho}, {Gorgoni}, {Green}, {Guy},
  {Harmer}, {Hernand ez}, {Honscheid}, {Huang}, {James}, {Jannuzi}, {Jiang},
  {Joyce}, {Karcher}, {Karkar}, {Kehoe}, {Kneib}, {Kueter-Young}, {Lan},
  {Lauer}, {Le Guillou}, {Le Van Suu}, {Lee}, {Lesser}, {Perreault Levasseur},
  {Li}, {Mann}, {Marshall}, {Mart{\'\i}nez-V{\'a}zquez}, {Martini}, {du Mas des
  Bourboux}, {McManus}, {Meier}, {M{\'e}nard}, {Metcalfe},
  {Mu{\~n}oz-Guti{\'e}rrez}, {Najita}, {Napier}, {Narayan}, {Newman}, {Nie},
  {Nord}, {Norman}, {Olsen}, {Paat}, {Palanque-Delabrouille}, {Peng},
  {Poppett}, {Poremba}, {Prakash}, {Rabinowitz}, {Raichoor}, {Rezaie},
  {Robertson}, {Roe}, {Ross}, {Ross}, {Rudnick}, {Safonova}, {Saha},
  {S{\'a}nchez}, {Savary}, {Schweiker}, {Scott}, {Seo}, {Shan}, {Silva},
  {Slepian}, {Soto}, {Sprayberry}, {Staten}, {Stillman}, {Stupak}, {Summers},
  {Sien Tie}, {Tirado}, {Vargas-Maga{\~n}a}, {Vivas}, {Wechsler}, {Williams},
  {Yang}, {Yang}, {Yapici}, {Zaritsky}, {Zenteno}, {Zhang}, {Zhang}, {Zhou}, \&
  {Zhou}}]{DR8_2019AJ....157..168D}
{Dey}, A., {Schlegel}, D.~J., {Lang}, D., {et~al.} 2019, \aj, 157, 168

\bibitem[{{Diaferio} {et~al.}(2001){Diaferio}, {Kauffmann}, {Balogh}, {White},
  {Schade}, \& {Ellingson}}]{2001Diaferio}
{Diaferio}, A., {Kauffmann}, G., {Balogh}, M.~L., {et~al.} 2001, \mnras, 323,
  999

\bibitem[{{Dressler}(1980)}]{dressler1980galaxy}
{Dressler}, A. 1980, \apj, 236, 351

\bibitem[{{Dressler}(1986)}]{Dressler1986_vdisp}
{Dressler}, A. 1986, \apj, 301, 35

\bibitem[{{Drinkwater} {et~al.}(2001{\natexlab{a}}){Drinkwater}, {Gregg}, \&
  {Colless}}]{drinkwater2001substructure}
{Drinkwater}, M.~J., {Gregg}, M.~D., \& {Colless}, M. 2001{\natexlab{a}},
  \apjl, 548, L139

\bibitem[{{Drinkwater} {et~al.}(2001{\natexlab{b}}){Drinkwater}, {Gregg},
  {Holman}, \& {Brown}}]{Drinkwater2001_evolutionofFornaxGal}
{Drinkwater}, M.~J., {Gregg}, M.~D., {Holman}, B.~A., \& {Brown}, M.~J.~I.
  2001{\natexlab{b}}, \mnras, 326, 1076

\bibitem[{{Ferguson}(1989)}]{FergusonFornax89}
{Ferguson}, H.~C. 1989, \aj, 98, 367

\bibitem[{{Frank} {et~al.}(2013){Frank}, {Peterson}, {Andersson}, {Fabian}, \&
  {Sanders}}]{2013ApJ...764...46F}
{Frank}, K.~A., {Peterson}, J.~R., {Andersson}, K., {Fabian}, A.~C., \&
  {Sanders}, J.~S. 2013, \apj, 764, 46

\bibitem[{{Gavazzi} {et~al.}(1999){Gavazzi}, {Boselli}, {Scodeggio}, {Pierini},
  \& {Belsole}}]{1999MNRAS.304..595G}
{Gavazzi}, G., {Boselli}, A., {Scodeggio}, M., {Pierini}, D., \& {Belsole}, E.
  1999, \mnras, 304, 595

\bibitem[{{Ger{\'e}b} {et~al.}(2016){Ger{\'e}b}, {Catinella}, {Cortese},
  {Bekki}, {Moran}, \& {Schiminovich}}]{2016MNRAS.462..382G}
{Ger{\'e}b}, K., {Catinella}, B., {Cortese}, L., {et~al.} 2016, \mnras, 462,
  382

\bibitem[{{Ger{\'e}b} {et~al.}(2018){Ger{\'e}b}, {Janowiecki}, {Catinella},
  {Cortese}, \& {Kilborn}}]{2018MNRAS.476..896G}
{Ger{\'e}b}, K., {Janowiecki}, S., {Catinella}, B., {Cortese}, L., \&
  {Kilborn}, V. 2018, \mnras, 476, 896

\bibitem[{{Giovanelli} \& {Haynes}(1983)}]{giovanelli1983AJ.....88..881G}
{Giovanelli}, R. \& {Haynes}, M.~P. 1983, \aj, 88, 881

\bibitem[{{Grillmair} {et~al.}(1994){Grillmair}, {Freeman}, {Bicknell},
  {Carter}, {Couch}, {Sommer-Larsen}, \& {Taylor}}]{Grillmair1994_ngc1399}
{Grillmair}, C.~J., {Freeman}, K.~C., {Bicknell}, G.~V., {et~al.} 1994, \apjl,
  422, L9

\bibitem[{{Gunn} \& {Gott}(1972)}]{1972ApJ...176....1G}
{Gunn}, J.~E. \& {Gott}, J.~Richard, I. 1972, \apj, 176, 1

\bibitem[{{Hamraz} {et~al.}(2019){Hamraz}, {Peletier}, {Khosroshahi},
  {Valentijn}, {den Brok}, \& {Venhola}}]{hamraz_scaling}
{Hamraz}, E., {Peletier}, R.~F., {Khosroshahi}, H.~G., {et~al.} 2019, \aap,
  625, A94

\bibitem[{{Haynes} \& {Giovanelli}(1984)}]{H&G84}
{Haynes}, M.~P. \& {Giovanelli}, R. 1984, \aj, 89, 758

\bibitem[{{Haynes} \& {Giovanelli}(1986)}]{Haynes1986ApJ...306..466H}
{Haynes}, M.~P. \& {Giovanelli}, R. 1986, \apj, 306, 466

\bibitem[{{Horellou} {et~al.}(2001){Horellou}, {Black}, {van Gorkom}, {Combes},
  {van der Hulst}, \& {Charmandaris}}]{Horellou2001_fornaxA}
{Horellou}, C., {Black}, J.~H., {van Gorkom}, J.~H., {et~al.} 2001, \aap, 376,
  837

\bibitem[{{Hubble} \& {Humason}(1931)}]{1931ApJ....74...43H}
{Hubble}, E. \& {Humason}, M.~L. 1931, \apj, 74, 43

\bibitem[{{Hughes} \& {Cortese}(2009)}]{HughesCortese2009_environm}
{Hughes}, T.~M. \& {Cortese}, L. 2009, \mnras, 396, L41

\bibitem[{{Iodice} {et~al.}(2016){Iodice}, {Capaccioli}, {Grado}, {Limatola},
  {Spavone}, {Napolitano}, {Paolillo}, {Peletier}, {Cantiello}, {Lisker},
  {Wittmann}, {Venhola}, {Hilker}, {D'Abrusco}, {Pota}, \&
  {Schipani}}]{FDSIodice2016}
{Iodice}, E., {Capaccioli}, M., {Grado}, A., {et~al.} 2016, \apj, 820, 42

\bibitem[{{Iodice} {et~al.}(2019{\natexlab{a}}){Iodice}, {Sarzi}, {Bittner},
  {Coccato}, {Costantin}, {Corsini}, {van de Ven}, {de Zeeuw},
  {Falc{\'o}n-Barroso}, {Gadotti}, {Lyubenova}, {Mart{\'\i}n-Navarro},
  {McDermid}, {Nedelchev}, {Pinna}, {Pizzella}, {Spavone}, \&
  {Viaene}}]{iodiceFDS2019}
{Iodice}, E., {Sarzi}, M., {Bittner}, A., {et~al.} 2019{\natexlab{a}}, \aap,
  627, A136

\bibitem[{{Iodice} {et~al.}(2017){Iodice}, {Spavone}, {Capaccioli}, {Peletier},
  {Richtler}, {Hilker}, {Mieske}, {Limatola}, {Grado}, {Napolitano},
  {Cantiello}, {D'Abrusco}, {Paolillo}, {Venhola}, {Lisker}, {Van de Ven},
  {Falcon-Barroso}, \& {Schipani}}]{FDSIodice_2017_fornaxA}
{Iodice}, E., {Spavone}, M., {Capaccioli}, M., {et~al.} 2017, \apj, 839, 21

\bibitem[{{Iodice} {et~al.}(2019{\natexlab{b}}){Iodice}, {Spavone},
  {Capaccioli}, {Peletier}, {van de Ven}, {Napolitano}, {Hilker}, {Mieske},
  {Smith}, {Pasquali}, {Limatola}, {Grado}, {Venhola}, {Cantiello}, {Paolillo},
  {Falcon-Barroso}, {D'Abrusco}, \& {Schipani}}]{2019A&A...623A...1I}
{Iodice}, E., {Spavone}, M., {Capaccioli}, M., {et~al.} 2019{\natexlab{b}},
  \aap, 623, A1

\bibitem[{{Jaff{\'e}} {et~al.}(2015){Jaff{\'e}}, {Smith}, {Candlish},
  {Poggianti}, {Sheen}, \& {Verheijen}}]{jaffecaustic}
{Jaff{\'e}}, Y.~L., {Smith}, R., {Candlish}, G.~N., {et~al.} 2015, \mnras, 448,
  1715

\bibitem[{{Jarrett} {et~al.}(2019){Jarrett}, {Cluver}, {Brown}, {Dale}, {Tsai},
  \& {Masci}}]{2019tom}
{Jarrett}, T.~H., {Cluver}, M.~E., {Brown}, M.~J.~I., {et~al.} 2019, \apjs,
  245, 25

\bibitem[{{Jarrett} {et~al.}(2013){Jarrett}, {Masci}, {Tsai}, {Petty},
  {Cluver}, {Assef}, {Benford}, {Blain}, {Bridge}, {Donoso}, {Eisenhardt},
  {Koribalski}, {Lake}, {Neill}, {Seibert}, {Sheth}, {Stanford}, \&
  {Wright}}]{Tom2013AJ}
{Jarrett}, T.~H., {Masci}, F., {Tsai}, C.~W., {et~al.} 2013, \aj, 145, 6

\bibitem[{{Jones} {et~al.}(2018){Jones}, {Espada}, {Verdes-Montenegro},
  {Huchtmeier}, {Lisenfeld}, {Leon}, {Sulentic}, {Sabater}, {Jones}, {Sanchez},
  \& {Garrido}}]{2018A&A...609A..17J}
{Jones}, M.~G., {Espada}, D., {Verdes-Montenegro}, L., {et~al.} 2018, \aap,
  609, A17

\bibitem[{{Jord{\'a}n} {et~al.}(2007){Jord{\'a}n}, {Blakeslee}, {C{\^o}t{\'e}},
  {Ferrarese}, {Infante}, {Mei}, {Merritt}, {Peng}, {Tonry}, \&
  {West}}]{JAndres2007_ACSfornax}
{Jord{\'a}n}, A., {Blakeslee}, J.~P., {C{\^o}t{\'e}}, P., {et~al.} 2007, \apjs,
  169, 213

\bibitem[{{Jorsater} \& {van Moorsel}(1995)}]{Jorsater1365}
{Jorsater}, S. \& {van Moorsel}, G.~A. 1995, \aj, 110, 2037

\bibitem[{{Kennicutt} \& {Evans}(2012)}]{kenni2012}
{Kennicutt}, R.~C. \& {Evans}, N.~J. 2012, \araa, 50, 531

\bibitem[{{Koribalski} {et~al.}(2004){Koribalski}, {Staveley-Smith}, {Kilborn},
  {Ryder}, {Kraan-Korteweg}, {Ryan-Weber}, {Ekers}, {Jerjen}, {Henning},
  {Putman}, {Zwaan}, {de Blok}, {Calabretta}, {Disney}, {Minchin}, {Bhathal},
  {Boyce}, {Drinkwater}, {Freeman}, {Gibson}, {Green}, {Haynes}, {Juraszek},
  {Kesteven}, {Knezek}, {Mader}, {Marquarding}, {Meyer}, {Mould}, {Oosterloo},
  {O'Brien}, {Price}, {Sadler}, {Schr{\"o}der}, {Stewart}, {Stootman}, {Waugh},
  {Warren}, {Webster}, \& {Wright}}]{BaerbelBGC2004}
{Koribalski}, B.~S., {Staveley-Smith}, L., {Kilborn}, V.~A., {et~al.} 2004,
  \aj, 128, 16

\bibitem[{{Kreckel} {et~al.}(2012){Kreckel}, {Platen}, {Arag{\'o}n-Calvo}, {van
  Gorkom}, {van de Weygaert}, {van der Hulst}, \& {Beygu}}]{VoidKreckel2012}
{Kreckel}, K., {Platen}, E., {Arag{\'o}n-Calvo}, M.~A., {et~al.} 2012, \aj,
  144, 16

\bibitem[{{Lauberts} \& {Valentijn}(1989)}]{1989spce.book.....L}
{Lauberts}, A. \& {Valentijn}, E.~A. 1989, {The surface photometry catalogue of
  the ESO-Uppsala galaxies}

\bibitem[{{Lee-Waddell} {et~al.}(2018){Lee-Waddell}, {Serra}, {Koribalski},
  {Venhola}, {Iodice}, {Catinella}, {Cortese}, {Peletier}, {Popping}, {Keenan},
  \& {Capaccioli}}]{karenNGC1427A}
{Lee-Waddell}, K., {Serra}, P., {Koribalski}, B., {et~al.} 2018, \mnras, 474,
  1108

\bibitem[{{Leroy} {et~al.}(2019){Leroy}, {Sandstrom}, {Lang}, {Lewis}, {Salim},
  {Behrens}, {Chastenet}, {Chiang}, {Gallagher}, {Kessler}, \&
  {Utomo}}]{Leroy2019ApJS}
{Leroy}, A.~K., {Sandstrom}, K.~M., {Lang}, D., {et~al.} 2019, \apjs, 244, 24

\bibitem[{{Lupton}(2005)}]{lupton2005}
{Lupton}, P. 2005

\bibitem[{{Maddox} {et~al.}(2015){Maddox}, {Hess}, {Obreschkow}, {Jarvis}, \&
  {Blyth}}]{Maddox2015}
{Maddox}, N., {Hess}, K.~M., {Obreschkow}, D., {Jarvis}, M.~J., \& {Blyth},
  S.~L. 2015, \mnras, 447, 1610

\bibitem[{{Maddox} {et~al.}(2019){Maddox}, {Serra}, {Venhola}, {Peletier},
  {Loubser}, \& {Iodice}}]{Maddox2019}
{Maddox}, N., {Serra}, P., {Venhola}, A., {et~al.} 2019, \mnras, 490, 1666

\bibitem[{{Marasco} {et~al.}(2016){Marasco}, {Crain}, {Schaye}, {Bah{\'e}},
  {van der Hulst}, {Theuns}, \& {Bower}}]{Antonino2016MNRAS.461.2630M}
{Marasco}, A., {Crain}, R.~A., {Schaye}, J., {et~al.} 2016, \mnras, 461, 2630

\bibitem[{{Matthews} {et~al.}(1998){Matthews}, {van Driel}, \&
  {Gallagher}}]{ESO015_1998AJ....116.1169M}
{Matthews}, L.~D., {van Driel}, W., \& {Gallagher}, J.~S., I. 1998, \aj, 116,
  1169

\bibitem[{{Meyer} {et~al.}(2017){Meyer}, {Robotham}, {Obreschkow}, {Westmeier},
  {Duffy}, \& {Staveley-Smith}}]{Meyertracing}
{Meyer}, M., {Robotham}, A., {Obreschkow}, D., {et~al.} 2017, \pasa, 34, 52

\bibitem[{{Mould} {et~al.}(2000){Mould}, {Hughes}, {Stetson}, {Gibson},
  {Huchra}, {Freedman}, {Kennicutt}, {Bresolin}, {Ferrarese}, {Ford}, {Graham},
  {Han}, {Hoessel}, {Illingworth}, {Kelson}, {Macri}, {Madore}, {Phelps},
  {Prosser}, {Rawson}, {Saha}, {Sakai}, {Sebo}, {Silbermann}, \&
  {Turner}}]{2000ApJ...528..655M}
{Mould}, J.~R., {Hughes}, S. M.~G., {Stetson}, P.~B., {et~al.} 2000, \apj, 528,
  655

\bibitem[{{Nasonova} {et~al.}(2011){Nasonova}, {de Freitas Pacheco}, \&
  {Karachentsev}}]{2011A&A...532A.104N}
{Nasonova}, O.~G., {de Freitas Pacheco}, J.~A., \& {Karachentsev}, I.~D. 2011,
  \aap, 532, A104

\bibitem[{{Navarro} {et~al.}(1996){Navarro}, {Frenk}, \& {White}}]{NFW}
{Navarro}, J.~F., {Frenk}, C.~S., \& {White}, S. D.~M. 1996, \apj, 462, 563

\bibitem[{{Oemler}(1974)}]{1974ApJ...194....1O}
{Oemler}, Augustus, J. 1974, \apj, 194, 1

\bibitem[{{Oh} {et~al.}(2018){Oh}, {Kim}, {Lee}, {Sheen}, {Kim}, {Ree}, {Ho},
  {Kyeong}, {Sung}, {Park}, \& {Yi}}]{2018ApJS..237...14O}
{Oh}, S., {Kim}, K., {Lee}, J.~H., {et~al.} 2018, \apjs, 237, 14

\bibitem[{{Ondrechen} \& {van der Hulst}(1989)}]{absorp1365}
{Ondrechen}, M.~P. \& {van der Hulst}, J.~M. 1989, \apj, 342, 29

\bibitem[{{Paolillo} {et~al.}(2002){Paolillo}, {Fabbiano}, {Peres}, \&
  {Kim}}]{paolillo2002xray}
{Paolillo}, M., {Fabbiano}, G., {Peres}, G., \& {Kim}, D.~W. 2002, \apj, 565,
  883

\bibitem[{{Peletier} {et~al.}(2020){Peletier}, {Iodice}, {Venhola},
  {Capaccioli}, {Cantiello}, {D'Abrusco}, {Falc{\'o}n-Barroso}, {Grado},
  {Hilker}, {Limatola}, {Mieske}, {Napolitano}, {Paolillo}, {Spavone},
  {Valentijn}, {van de Ven}, \& {Verdoes Kleijn}}]{Reynier_2020arXiv200812633P}
{Peletier}, R., {Iodice}, E., {Venhola}, A., {et~al.} 2020, arXiv e-prints,
  arXiv:2008.12633

\bibitem[{{Raj} {et~al.}(2019){Raj}, {Iodice}, {Napolitano}, {Spavone}, {Su},
  {Peletier}, {Davis}, {Zabel}, {Hilker}, {Mieske}, {Falcon Barroso},
  {Cantiello}, {van de Ven}, {Watkins}, {Salo}, {Schipani}, {Capaccioli}, \&
  {Venhola}}]{FDS_RAJ}
{Raj}, M.~A., {Iodice}, E., {Napolitano}, N.~R., {et~al.} 2019, \aap, 628, A4

\bibitem[{{Rhee} {et~al.}(2017){Rhee}, {Smith}, {Choi}, {Yi}, {Jaff{\'e}},
  {Candlish}, \& {S{\'a}nchez-J{\'a}nssen}}]{rhee_2017ApJ}
{Rhee}, J., {Smith}, R., {Choi}, H., {et~al.} 2017, \apj, 843, 128

\bibitem[{{Rodr{\'\i}guez-Ardila} {et~al.}(2017){Rodr{\'\i}guez-Ardila},
  {Prieto}, {Mazzalay}, {Fern{\'a}ndez-Ontiveros}, {Luque}, \&
  {M{\"u}ller-S{\'a}nchez}}]{2017MNRAS.470.2845R}
{Rodr{\'\i}guez-Ardila}, A., {Prieto}, M.~A., {Mazzalay}, X., {et~al.} 2017,
  \mnras, 470, 2845

\bibitem[{{Sault} {et~al.}(1995){Sault}, {Teuben}, \& {Wright}}]{SaultMiriad}
{Sault}, R.~J., {Teuben}, P.~J., \& {Wright}, M.~C.~H. 1995, Astronomical
  Society of the Pacific Conference Series, Vol.~77, {A Retrospective View of
  MIRIAD}, ed. R.~A. {Shaw}, H.~E. {Payne}, \& J.~J.~E. {Hayes}, 433

\bibitem[{{Schr{\"o}der} {et~al.}(2001){Schr{\"o}der}, {Drinkwater}, \&
  {Richter}}]{Schroder2001}
{Schr{\"o}der}, A., {Drinkwater}, M.~J., \& {Richter}, O.~G. 2001, \aap, 376,
  98

\bibitem[{{Schweizer}(1980)}]{Schweizer_opticalFornaxA}
{Schweizer}, F. 1980, \apj, 237, 303

\bibitem[{{Seibert} {et~al.}(2012){Seibert}, {Wyder}, {Neill}, {Madore},
  {Bianchi}, {Smith}, {Shiao}, {Schiminovich}, {Rich}, {Conrow}, {Martin}, \&
  {GALEX Catalog Team}}]{GCAT2012}
{Seibert}, M., {Wyder}, T., {Neill}, J., {et~al.} 2012, in American
  Astronomical Society Meeting Abstracts, Vol. 219, American Astronomical
  Society Meeting Abstracts \#219, 340.01

\bibitem[{{Serra} {et~al.}(2016){Serra}, {de Blok}, {Bryan}, {Colafrancesco},
  {Dettmar}, {Frank}, {Govoni}, {Jozsa}, {Kraan-Korteweg}, {Maccagni},
  {Loubser}, {Murgia}, {Oosterloo}, {Peletier}, {Pizzo}, {Richter},
  {Ramatsoku}, {Smith}, {Trager}, {van Gorkom}, \&
  {Verheijen}}]{2016PaoloMeerkat}
{Serra}, P., {de Blok}, W.~J.~G., {Bryan}, G.~L., {et~al.} 2016, in MeerKAT
  Science: On the Pathway to the SKA, 8

\bibitem[{{Serra} {et~al.}(2012){Serra}, {Jurek}, \&
  {Fl{\"o}er}}]{PaoloReliability}
{Serra}, P., {Jurek}, R., \& {Fl{\"o}er}, L. 2012, \pasa, 29, 296

\bibitem[{{Serra} {et~al.}(2019){Serra}, {Maccagni}, {Kleiner}, {de Blok}, {van
  Gorkom}, {Hugo}, {Iodice}, {J{\'o}zsa}, {Kamphuis}, {Kraan-Korteweg}, {Loni},
  {Makhathini}, {Moln{\'a}r}, {Oosterloo}, {Peletier}, {Ramaila}, {Ramatsoku},
  {Smirnov}, {Smith}, {Spavone}, {Thorat}, {Trager}, \&
  {Venhola}}]{Paolo2019_fornaxA}
{Serra}, P., {Maccagni}, F.~M., {Kleiner}, D., {et~al.} 2019, \aap, 628, A122

\bibitem[{{Serra} {et~al.}(2015){Serra}, {Westmeier}, {Giese}, {Jurek},
  {Fl{\"o}er}, {Popping}, {Winkel}, {van der Hulst}, {Meyer}, {Koribalski},
  {Staveley-Smith}, \& {Courtois}}]{PaoloSoFiA}
{Serra}, P., {Westmeier}, T., {Giese}, N., {et~al.} 2015, \mnras, 448, 1922

\bibitem[{{Sheardown} {et~al.}(2018){Sheardown}, {Roediger}, {Su}, {Kraft},
  {Fish}, {ZuHone}, {Forman}, {Jones}, {Churazov}, \&
  {Nulsen}}]{Sheardown_Fornaxhistory}
{Sheardown}, A., {Roediger}, E., {Su}, Y., {et~al.} 2018, \apj, 865, 118

\bibitem[{{Sheen} {et~al.}(2012){Sheen}, {Yi}, {Ree}, \&
  {Lee}}]{2012ApJS..202....8S}
{Sheen}, Y.-K., {Yi}, S.~K., {Ree}, C.~H., \& {Lee}, J. 2012, \apjs, 202, 8

\bibitem[{{Solanes} {et~al.}(1996){Solanes}, {Giovanelli}, \&
  {Haynes}}]{SolanesHImassfunc}
{Solanes}, J.~M., {Giovanelli}, R., \& {Haynes}, M.~P. 1996, \apj, 461, 609

\bibitem[{{Solanes} {et~al.}(2001){Solanes}, {Manrique},
  {Garc{\'\i}a-G{\'o}mez}, {Gonz{\'a}lez-Casado}, {Giovanelli}, \&
  {Haynes}}]{Solanes2001_DEF}
{Solanes}, J.~M., {Manrique}, A., {Garc{\'\i}a-G{\'o}mez}, C., {et~al.} 2001,
  \apj, 548, 97

\bibitem[{{Spavone} {et~al.}(2020){Spavone}, {Iodice}, {van de Ven},
  {Falc{\'o}n-Barroso}, {Raj}, {Hilker}, {Peletier}, {Capaccioli}, {Mieske},
  {Venhola}, {Napolitano}, {Cantiello}, {Paolillo}, \&
  {Schipani}}]{2020A&A...639A..14S}
{Spavone}, M., {Iodice}, E., {van de Ven}, G., {et~al.} 2020, \aap, 639, A14

\bibitem[{{Spiniello} {et~al.}(2018){Spiniello}, {Napolitano}, {Arnaboldi},
  {Tortora}, {Coccato}, {Capaccioli}, {Gerhard}, {Iodice}, {Spavone},
  {Cantiello}, {Peletier}, {Paolillo}, \& {Schipani}}]{2018MNRAS.477.1880S}
{Spiniello}, C., {Napolitano}, N.~R., {Arnaboldi}, M., {et~al.} 2018, \mnras,
  477, 1880

\bibitem[{{Su} {et~al.}(2021){Su}, {Salo}, {Janz}, {Laurikainen}, {Venhola},
  {Peletier}, {Iodice}, {Hilker}, {Cantiello}, {Napolitano}, {Spavone}, {Raj},
  {van de Ven}, {Mieske}, {Paolillo}, {Capaccioli}, {Valentijn}, \&
  {Watkins}}]{2021arXiv210105699S}
{Su}, A.~H., {Salo}, H., {Janz}, J., {et~al.} 2021, arXiv e-prints,
  arXiv:2101.05699

\bibitem[{{Theureau} {et~al.}(1998){Theureau}, {Bottinelli}, {Coudreau-Durand},
  {Gouguenheim}, {Hallet}, {Loulergue}, {Paturel}, \&
  {Teerikorpi}}]{ESO051_1998A&AS..130..333T}
{Theureau}, G., {Bottinelli}, L., {Coudreau-Durand}, N., {et~al.} 1998, \aaps,
  130, 333

\bibitem[{{Toomre} \& {Toomre}(1972)}]{1972ApJ...178..623T}
{Toomre}, A. \& {Toomre}, J. 1972, \apj, 178, 623

\bibitem[{{van der Hulst} {et~al.}(1983){van der Hulst}, {Ondrechen}, {van
  Gorkom}, \& {Hummel}}]{vanderHulst1365}
{van der Hulst}, J.~M., {Ondrechen}, M.~P., {van Gorkom}, J.~H., \& {Hummel},
  E. 1983, in IAU Symposium, Vol. 100, Internal Kinematics and Dynamics of
  Galaxies, ed. E.~{Athanassoula}, 233--234

\bibitem[{{Venhola} {et~al.}(2018){Venhola}, {Peletier}, {Laurikainen}, {Salo},
  {Iodice}, {Mieske}, {Hilker}, {Wittmann}, {Lisker}, {Paolillo}, {Cantiello},
  {Janz}, {Spavone}, {D'Abrusco}, {Ven}, {Napolitano}, {Kleijn}, {Maddox},
  {Capaccioli}, {Grado}, {Valentijn}, {Falc{\'o}n-Barroso}, \&
  {Limatola}}]{venhola2018fds}
{Venhola}, A., {Peletier}, R., {Laurikainen}, E., {et~al.} 2018, \aap, 620,
  A165

\bibitem[{{Venhola} {et~al.}(2019){Venhola}, {Peletier}, {Laurikainen}, {Salo},
  {Iodice}, {Mieske}, {Hilker}, {Wittmann}, {Paolillo}, {Cantiello}, {Janz},
  {Spavone}, {D'Abrusco}, {van de Ven}, {Napolitano}, {Verdoes Kleijn},
  {Capaccioli}, {Grado}, {Valentijn}, {Falc{\'o}n-Barroso}, \&
  {Limatola}}]{venhola2019A&A}
{Venhola}, A., {Peletier}, R., {Laurikainen}, E., {et~al.} 2019, \aap, 625,
  A143

\bibitem[{{Wang} {et~al.}(2016){Wang}, {Koribalski}, {Serra}, {van der Hulst},
  {Roychowdhury}, {Kamphuis}, \& {Chengalur}}]{wang2016_size}
{Wang}, J., {Koribalski}, B.~S., {Serra}, P., {et~al.} 2016, \mnras, 460, 2143

\bibitem[{{Waugh}(2005)}]{waughphd}
{Waugh}, M. 2005, PhD thesis, University of Melbourne.

\bibitem[{{Waugh} {et~al.}(2002){Waugh}, {Drinkwater}, {Webster},
  {Staveley-Smith}, {Kilborn}, {Barnes}, {Bhathal}, {de Blok}, {Boyce},
  {Disney}, {Ekers}, {Freeman}, {Gibson}, {Henning}, {Jerjen}, {Knezek},
  {Koribalski}, {Marquarding}, {Minchin}, {Price}, {Putman}, {Ryder}, {Sadler},
  {Stootman}, \& {Zwaan}}]{Waugh2002}
{Waugh}, M., {Drinkwater}, M.~J., {Webster}, R.~L., {et~al.} 2002, \mnras, 337,
  641

\bibitem[{{Whitaker} {et~al.}(2012){Whitaker}, {van Dokkum}, {Brammer}, \&
  {Franx}}]{Whitaker2012ApJ}
{Whitaker}, K.~E., {van Dokkum}, P.~G., {Brammer}, G., \& {Franx}, M. 2012,
  \apjl, 754, L29

\bibitem[{{Zabel} {et~al.}(2020){Zabel}, {Davis}, {Sarzi}, {Nedelchev},
  {Chevance}, {Kruijssen}, {Iodice}, {Baes}, {Bendo}, {Corsini}, {De Looze},
  {de Zeeuw}, {Gadotti}, {Grossi}, {Peletier}, {Pinna}, {Serra}, {van de
  Voort}, {Venhola}, {Viaene}, \& {Vlahakis}}]{zabel2}
{Zabel}, N., {Davis}, T.~A., {Sarzi}, M., {et~al.} 2020, \mnras

\bibitem[{{Zabel} {et~al.}(2019){Zabel}, {Davis}, {Smith}, {Maddox}, {Bendo},
  {Peletier}, {Iodice}, {Venhola}, {Baes}, {Davies}, {de Looze}, {Gomez},
  {Grossi}, {Kenney}, {Serra}, {van de Voort}, {Vlahakis}, \& {Young}}]{zabel}
{Zabel}, N., {Davis}, T.~A., {Smith}, M. W.~L., {et~al.} 2019, \mnras, 483,
  2251

\end{thebibliography}

\twocolumn
\appendix
\section{}
\label{sec:app}
    
    In this section, we describe in detail each galaxy of our sample. Whenever possible we compare the information estimated from our \hi{} data with observations at other wavelengths. The galaxies are sorted according to increasing \hi{} mass as in Table~\ref{tab:mass}, Fig.~\ref{fig:allmorph} and Fig.~\ref{fig:allprof}.  
    
    \subsection*{FCC~323}
    According to \cite{FergusonFornax89}, FCC~323 is a dE2/ImV.
    This is the faintest among our \hi{} detections and, for the first time, we detected it in \hi{} and thus measured its recessional velocity of \vbary{}~$=$~1502 \kms{}. FCC~323 has a \mhi{}~=~(7.6$\pm$3.9)$\times$10$^6$~M$_\odot$ and a \mst{}~=~(6.3$\pm$5.8)$\times$10$^7$~M$_\odot$.
    The Fornax environment might have strongly affected the \hi{} reservoir of FCC~323. It has the highest \mhi{}/\mst{} offset from the (extrapolated) xGASS scaling relation in Fig.~\ref{fig:himfraction}.
    Its closest galaxy both on the sky and velocity is NGC~1437B (120~kpc and 12~\kms{} away), 
    leading us to think that FCC~323 and NGC~1437B might be part of a subgroup or a pair of galaxies.
    
    \subsection*{FCC~120 }
    FCC~120 is classified as ImIV by \cite{FergusonFornax89}.
    The \HI{} emission associated with the galaxy seems to be quite regular.
    The ATCA spectrum in Fig.~\ref{fig:allprof} shows one peak (within the uncertainties), which is different from the 100 - \kms{} - wide double horn spectrum detected by \cite{Schroder2001} based on Parkes data.
    Even if the latter data shows strong RFI at a velocity of 1250\kms{}, this RFI does not affect the detection of the galaxy. 
    However, \vspe{} from \cite{Maddox2019}, agrees with our \vbary{}. 
    Deeper \hi{} data (\citealt{2016PaoloMeerkat}) may confirm this. 
    FCC~120 has a \mhi{}~=~(4.5$\pm$1.5)$\times$10$^7$~M$_\odot$ and a \mst{}~=~(1.6$\pm$0.5)$\times$10$^8$~M$_\odot$.
    
    \subsection*{FCC~102}
    FCC~102 is an irregular ImIV classified by \cite{FergusonFornax89}. 
    We detect \hi{} emission associated with FCC~102 for the first time. Although the \hi{} morphology is not resolved, the \hi{} is offset towards the north with respect to the optical centre. The difference between \vspe{} from \cite{Maddox2019}, and \vbary{} from ATCA data is almost 100 \kms{}. However, those values are still consistent due to the large uncertainty on \vspe{}, which was measured from absorption lines for this galaxy (Natasha Maddox, priv. comm.). Recently this galaxy was detected during MeerKAT test observations in preparation for the MeerKAT Fornax Survey \cite{2016PaoloMeerkat}, confirming our ATCA \hi{} results. FCC~102 has a \mhi{}~=~(4.6$\pm$1.6)$\times$10$^7$~M$_\odot$ and a \mst{}~=~(1.6$\pm$0.5)$\times$10$^8$~M$_\odot$. The comparison with \hi{} belonging to non-cluster galaxies with similar \mst{} in Fig.~\ref{fig:himfraction} shows that FCC~102 is \hi{} deficient.
    
    We identify FCC~102 to be part of the NGC~1365 infalling subgroup. It lies between NGC~1365 and FCC~090 both on the sky and in velocity (see Fig.~\ref{fig:distrib} and Fig.~\ref{fig:phsp}).
    Furthermore the \hi{} elongation of FCC~102 is in the same direction as the \hi{} elongation of NGC~1365. 
     `
    \subsection*{NGC~1436}
    NGC~1436 is the closest \hi{} detected spiral to the centre of the cluster (see Fig~\ref{fig:distrib} and Fig.~\ref{fig:phsp}). \cite{FergusonFornax89} classify this galaxy as ScII.
    The \hi{} emission associated with NGC~1436 is confined well within the stellar body suggesting a truncated \hi{} disc.
    However, Fig.~\ref{fig:allprof} shows that the ATCA spectrum compared with that based on GBT data \citep{courtoisGBT} misses flux from the blushifted part of the disc. This discrepancy might be due to a faint \hi{} emission of NGC~1436 which extends over a region smaller than the ATCA beam in at least some of the blue-shifted channels. This would cause the average HI column density within the beam to fall below the sensitivity limit, resulting in a loss of \hi{} flux in our data.
    The \hi{} spectrum of NGC~1436 spectrum from \cite{Schroder2001} is more similar to that obtained with the GBT than to ours.
    The \mhi{} estimated by the ATCA flux is (5.8$\pm$2.5)$\times$10$^7$~M$_\odot$, while the \mst{} of the galaxy is (1.6$\pm$0.4)$\times$10$^{10}$~M$_\odot$.
    Whether adopting the ATCA or the GBT \hi{} flux measurement, NGC~1436 would still be an \hi{} deficient galaxy (see Sect.\ref{subse:himfrac}). Similarly, it would still be the (\hi{} and \htwo{0pt}) detection with the largest \mhtwo{}/\mhi{} ratio in Fig.~\ref{fig:hionh2}.
    
    The high offset from the scaling relation in Fig.~\ref{fig:hionh2}, the truncated \hi{} disc in Fig.~\ref{fig:allmorph} and the molecular spiral structure detected by \cite{zabel} show that NGC~1436 may have gone through a quick interaction with the cluster environment, which did not affect the inner spiral structure yet.
    
    Furthermore, \cite{FDS_RAJ} found that NGC~1436 appears to be transforming into a S0 in its morphological evolution. They found spiral arms only in the central region whereas its outer disc resembles the smooth structure typically found in the discs of S0 galaxies.  
    The truncated \hi{} disc agrees with the observation of a passively-evolving outer optical disc, which is being shaped by the Fornax environment. 
     
    \subsection*{FCC~090}
    FCC~090 is classified as a peculiar elliptical galaxy by \cite{FergusonFornax89}.
    We detected \hi{} emission in FCC~090 for the first time. The \hi{} morphology of FCC~090 is elongated towards the south.
    FCC~090 has a \mhi{}~=~(5.9$\pm$2.1)$\times$10$^7$~M$_\odot$ and a \mst{}~=~(1.3$\pm$0.3)$\times$10$^9$~M$_\odot$.
    Its \mhi{}/\mst{} ratio makes it a \hi{} deficient galaxy, which lies on the edge of comparison sample of non-cluster galaxies (Fig.~\ref{fig:himfraction}).

    FCC~090 is part of the NGC~1365 infalling subgroup. Within the subgroup, the \hi{} morphology of FCC~090 is the only one which is extended towards the south; it is the farther galaxy from NGC~1365; it has the highest velocity in phase space. 
    FCC~090 is also the westernmost galaxy in the triplet of the aligned \hi{} disturbed galaxies. Besides FCC~090, this triplet includes FCC~102 and NGC~1365.
    
    The \hi{} and molecular distributions (\citealt{zabel}) are difficult to compare due to the different resolutions between ATCA and ALMA. 
    The molecular distribution extends beyond the stellar body showing a tail that points to the west in projection but it is still within our \hi{} detection that does not show any peculiarity in that direction.
    Our knowledge on \hi{} in FCC~090 as well as its belonging to the NGC~1365 subgroup, corroborates the idea of an infalling galaxy which is loosing the external gas envelope. 
    
    In the sample of dwarf galaxies studied by \cite{hamraz_scaling} FCC~090 is one out of two outliers in the Fornax cluster with a very blue inner part. 
    Recently, \cite{zabel2} showed that depletion time is shorter than usual (0.5~Gyr rather than 2~Gyr) which may be a consequence of the environmental interactions taking place in it. 
     
     \subsection*{ESO~358-G016}
     ESO~358-G016 is classified as Sdm (edge-on) by \cite{FergusonFornax89}. It has a \mhi{}~=~(7.4$\pm$2.5)$\times$10$^7$~M$_\odot$ and a \mst{}~=~(1.2$\pm$0.7)$\times$10$^8$~M$_\odot$. ESO~358-G016 is the only member of the NGC~1365 infalling subgroup with a regular \hi{} morphology. Like all the low mass members of the subgroup, it is a \hi{} deficient galaxy. 
     It is also the second closest galaxy to NGC~1365 both on the sky and in phase space.
     \citealt{FDS_RAJ} suggested that ESO~358-G016 might have experienced disruptions of the outer disc due to the gravitational potential of the cluster. If that is the case, some \hi{} survived in this process. 
   
     \subsection*{ESO~358-G015}
     ESO~358-G015 is classified as a peculiar Scd-III by \cite{FergusonFornax89}.
     It has a regular \hi{} morphology, with a systemic velocity close to the recessional velocity of the cluster. It has a \mhi{}~=~(9.3$\pm$2.8)$\times$10$^7$~M$_\odot$ and a \mst{}~=~(7.6$\pm$2.1)$\times$10$^8$~M$_\odot$. Fig.~\ref{fig:himfraction} shows that ESO~358-G015 is a \hi{} deficient galaxy. 
     Furthermore, it is one of the four \hi{} detected galaxies that are north of NGC~1399.
     \cite{FDS_RAJ} identified a lopsided tail pointing to the centre of the cluster. They also suggested ESO~358-G015 to be a galaxy that is being pulled into the cluster centre, in the southern direction. The quite regular \hi{} morphology does not provide clear support to this hypothesis.

    \subsection*{ESO~358-G051}
    ESO~358-G051 is classified as SBcd-III by \cite{FergusonFornax89}. 
    The \hi{} distribution is regular and peaks on the optical centre. It is located north of NGC~1399 and it is the closest \hi{} deficient galaxy from it ($\sim$0.4~$R_\mathrm{vir}$). In general, among all our \hi{} detections, only NGC~1427A is closer to NGC~1399.
    ESO~358-G051 has a \mhi{}~=~(1.0$\pm$0.3)$\times$10$^8$~M$_\odot$ and a \mst{}~=~(2.1$\pm$0.5)$\times$10$^9$~M$_\odot$. Fig.~\ref{fig:himfraction} shows that ESO~358-G051 is a \hi{} deficient galaxy. 
    
    \cite{FDS_RAJ} marked this galaxy as a recent infaller with strong central H$\alpha$ emission powered by star formation. \cite{iodiceFDS2019} observed an ionised gas distribution extended towards the north. 
    This is the opposite direction with respect to the \htwo{0pt} elongation detected by \cite{zabel}.
    However, the \hi{} distribution does not seem to be disturbed, perhaps due to the ATCA resolution. 
    Although ESO~358-G051 is a \hi{} deficient galaxy, the unresolved \hi{} morphology does not provide information about on going interaction with the Fornax environment.

      \subsection*{ESO-LV 3580611}
      ESO-LV~3580611 is classified as SBm-III by \cite{FergusonFornax89}.
      The \hi{} morphology shows a lopsided \hi{} distribution in the northern part of the system. However, optical and \hi{} distribution share the same centre. 
      It has a \mhi{}~=~(1.1$\pm$0.3)$\times$10$^8$~M$_\odot$ and a \mst{}~=~(1.0$\pm$0.4)$\times$10$^8$~M$_\odot$.
      ESO-LV~3580611 is also one of the two \hi{} deficient galaxies in our sample that is outside the caustic curves (Fig.~\ref{fig:phsp}). \cite{Schroder2001}, based on the high deviation of ESO-LV~3580611 from the Tully-Fisher relation, suggested that this galaxy is falling into the cluster from the background.
      
      According to \cite{FDS_RAJ}, ESO-LV~3580611 may have experienced disruptions in the outskirts of the disc due to gravitational potential well of the cluster. Thus, similarly to ESO~358-G016, \hi{} has survived during the infall.

      \subsection*{NGC~1437B}
      NGC~1437B is classified as Sd (edge-on) by \cite{FergusonFornax89}. The \hi{} distribution shows an elongation to the south, while the \hi{} peak corresponds to the optical centroid (Fig.~\ref{fig:allmorph}).
      
      This galaxy is located on the south east of the cluster, close to FCC~323 both on the sky and in velocity (see Fig.~\ref{fig:distrib} and Fig.~\ref{fig:phsp}). Given the small separation of only 120~kpc with a difference of $\sim$10\kms{} in velocity, they might be part of an interacting subgroup of galaxies. NGC~1437B has a \mhi{}~=~(2.4$\pm$0.6)$\times$10$^8$~M$_\odot$ and a \mst{}~=~(5.00$\pm$0.01)$\times$10$^9$~M$_\odot$ and it is a \hi{} rich galaxy.
      
      Fig.~\ref{fig:hionh2} shows that the Fornax environment has not strongly affected the \mhtwo{}/\mhi{} ratio yet.  
      On the other hand, contrary to the case of NGC~1436, the two gas phases are probably experiencing the same environmental effect.
      Indeed, although the scales are different, we found consistency between the southern \hi{} elongation of NGC~1437B and the elongation of the molecular morphology detected by \cite{zabel}. Thus, the agreement between its \mhtwo{}/\mhi{} ratio and the xGASS scaling relation might be due to a slower evolution with respect to that taking place in NGC~1436.
      
      \cite{FDS_RAJ} did not observed any hint of optical morphological transition from the late-type spiral structure, which supports the hypothesis of a recent infaller. However, they found an optical disturbance in form of tidal tail, which points to the south in projection and might be due to a recent fly-by of another galaxy in the cluster. The \hi{} morphology agrees with the optical detected tail. If FCC~323 and NGC~1437B belong to the same subgroup, the former may be the fly-by galaxy that NGC~1437B has interacted with.

     \subsection*{NGC~1351A}
     Classified as Sc (edge-on) by \cite{FergusonFornax89}. The \hi{} distribution shows a projected elongation towards the south.
     It has a velocity similar to the recessional of the cluster (Fig.~\ref{fig:phsp}).
     NGC~1351A has a \mhi{}~=~(4.7$\pm$1.2)$\times$10$^8$~M$_\odot$ and a \mst{}~=~(3.5$\pm$0.8)$\times$10$^9$~M$_\odot$.
     The \hi{} reservoir is similar to that of non-cluster galaxies with the same \mst{} (Fig.~\ref{fig:himfraction}).
     NGC~1351A is isolated both on the sky and in  projected phase space (Fig.~ \ref{fig:distrib} and Fig.~\ref{fig:phsp}). Thus, the \hi{} asymmetry might be due to the interaction with ICM. \cite{zabel} detected a slightly more diffuse east side in the \htwo{0pt} distribution.
     
     \subsection*{NGC~1437A}
     NGC~1437A is classified as SdIII by \cite{FergusonFornax89}.
     The \hi{} morphology is quite regular and its \hi{} content is comparable with that of non-cluster galaxies with the same \mst{} (see Fig.s \ref{fig:allmorph} and \ref{fig:himfraction}).
     NGC~1437A has a \mhi{}~=~(5.6$\pm$1.3)$\times$10$^8$~M$_\odot$ and a \mst{}~=~(1.0$\pm$0.2)$\times$10$^9$~M$_\odot$.
     Although the optical appearance of NGC~1437A is not regular and the location of its star forming regions suggests that the galaxy is travelling in a south-east direction (\citealt{FDS_RAJ}), the \hi{} distribution does not show any strong asymmetries. Thus, ram-pressure and tidal interaction might not be the cause of the asymmetric star forming regions. However, the poor resolution of the \hi{} image may hide \hi{} asymmetries in projection.
     
\subsection*{ESO~358-G060}
ESO~358-G060 is a low mass galaxy classified as Sdm (edge-on) by \cite{FergusonFornax89}.
The \hi{} morphology is regular. ESO~358-G060 has a \mhi{}~=~(1.1$\pm$0.2)$\times$10$^9$~M$_\odot$ and a \mst{}~=~(1.0$\pm$0.6)$\times$10$^8$~M$_\odot$. Fig.~\ref{fig:himfraction} shows that ESO~358-G060 has a \hi{} reservoir comparable with that of non-cluster galaxies with the same \mst{}. As discussed in Sect.\ref{subse:himfrac}, we pointed out that the Fornax cluster environment is more effective in altering the gas content for galaxies with \mst{}~$<$~3$\times$10$^9$~M$_\odot$. This supports the idea that ESO~358-G060 is a likely new Fornax member which has not been affected by the cluster environment yet.
\cite{FDS_RAJ} observed irregular star-forming regions making the hypothesis of a disruptions due to the gravitational potential well of the cluster centre, during the fall. 
However, since the \hi{} distribution is not perturbed yet by environmental interactions, internal feedback might be the cause of the irregular star forming regions. Furthermore no \htwo{0pt} was detected by \cite{zabel} and it is a dust poor galaxy as discussed in Sect.\ref{sec:disc}. Overall, this galaxy is forming star with lower rate than that predicted for galaxies with similar \mhi{}/\mst{} (Fig.~\ref{fig:deltasfr}).

    \subsection*{ESO~358-G063}
    Classified as Scd (edge-on) by \cite{FergusonFornax89}.
    The atomic hydrogen distribution is more extended to the south-east part of the system. On this side, \hi{} contours are more spaced with respect to the opposite side of the disc where \hi{} emission seems to be confined within the stellar body.
    ESO~358-G063 has a \mhi{}~=~(1.7$\pm$0.4)$\times$10$^9$~M$_\odot$ and a \mst{}~=~(1.1$\pm$0.2)$\times$10$^{10}$~M$_\odot$.
    Despite the asymmetries in the \hi{} distribution, ESO~358G-063 is not \hi{} deficient, and the molecular gas detected by \cite{zabel} has a regular morphology. Thus, the galaxy has just started to interact with the cluster environment. The idea is supported by its position in projected phase space (Fig.~\ref{fig:phsp}) where the galaxy is just outside the caustic curves.
    \cite{FDS_RAJ} found irregular star-forming regions in the ill-defined spiral arms of ESO~358-G063, which may be signs of minor mergers.

      \subsection*{NGC~1427A}
      NGC~1427A is classified as Im-III by \cite{FergusonFornax89}.
      The \hi{} morphology shows a very long tail pointing to the south-east in projection (opposite direction to the centre of the cluster - we refer the reader to \cite{karenNGC1427A} for a detailed study about the origin of the tail).
      NGC~1427A has a \mhi{}~=~(2.1$\pm$0.4)$\times$10$^9$~M$_\odot$ and a \mst{}~=~(2.3$\pm$0.5)$\times$10$^9$~M$_\odot$.
      This is the second most massive \hi{} galaxy of our sample, which has an \hi{} content comparable to that of non-cluster (see Fig.~\ref{fig:himfraction}). 
      NGC~1427A is the closest \hi{} detection to the centre of the cluster (0.2~$R_\mathrm{vir}$).
      Due to its high velocity (see Fig.~\ref{fig:phsp}) it is not clear whether NGC~1427A is already virialised in the cluster or not. If it is virialised, it may decrease its velocity while reaching its apocentre.
      It is the only \mhtwo{}/\mhi{} upper limit with \mst{}~$>3\times$10$^9$~M$_\odot$ in Fig.~\ref{fig:hionh2}. Despite the large \hi{} reservoir no \htwo{0pt} was detected by \cite{zabel}. This may be due to a recent merger which involved the NGC~1427A progenitors that might have lowered its metallicity. 
      The \hi{}-to-\htwo{0pt} conversion might be inefficient to have a SFR consistent with its \mhi{}/\mst{} ratio (Fig.~\ref{fig:deltasfr}).

         \subsection*{NGC~1365}
         
         NGC~1365, the large barred spiral galaxy (SBbc(s)I by \citealt{FergusonFornax89}), is the most massive galaxy in our sample. The optical morphology shows a well defined spiral structure, three northern arms and two more compressed southern ones. 
         The \hi{} morphology (Fig.~\ref{fig:allmorph}) shows a very prominent extension to the north of the system and more dense contours corresponding to the two compressed optical arms.
         A very detailed study on the structure of NGC~1365 is made by \cite{Jorsater1365}. They also suggested that the motion of the galaxy through the ICM can account for its unwinding arms.
         
        NGC~1365 has a \mhi{}~=~(1.5$\pm$0.3)$\times$10$^{10}$~M$_\odot$ and a \mst{}~=~(6.2$\pm$1.3)$\times$10$^{10}$~M$_\odot$.
        Fig.~\ref{fig:himfraction} shows that NGC~1365 is the only galaxy in our sample which is at the upper edge of the comparison sample of non-cluster galaxies due to its high \mhi{}/\mst{}. Besides evidence of environmental interactions from its morphology, the strong gravitational potential appears to have been able to hold a sufficient amount of \hi{} to make it comparable with the general behaviour of local galaxies (see also Fig.~\ref{fig:hionh2}).
        
        NGC~1365 is the main member of the detected infalling subgroup of galaxies located to the south-west of the cluster. It is also the only galaxy in the subgroup with a normal \hi{} content.
        The existence of the NGC~1365 subgroup is also supported by \cite{drinkwater2001substructure} which found NGC~1365 to be part of a substructure. 
         
        There is a \hi{} hole in the centre of the system where \cite{zabel} detected warped \htwo{0pt} emission due to the central bar.
        \hi{} absorption was also investigated in \cite{absorp1365}.

\onecolumn

\section{}\label{sec:app2}
    \begin{figure*}[h!]
        \centering
        \includegraphics[width=170mm]{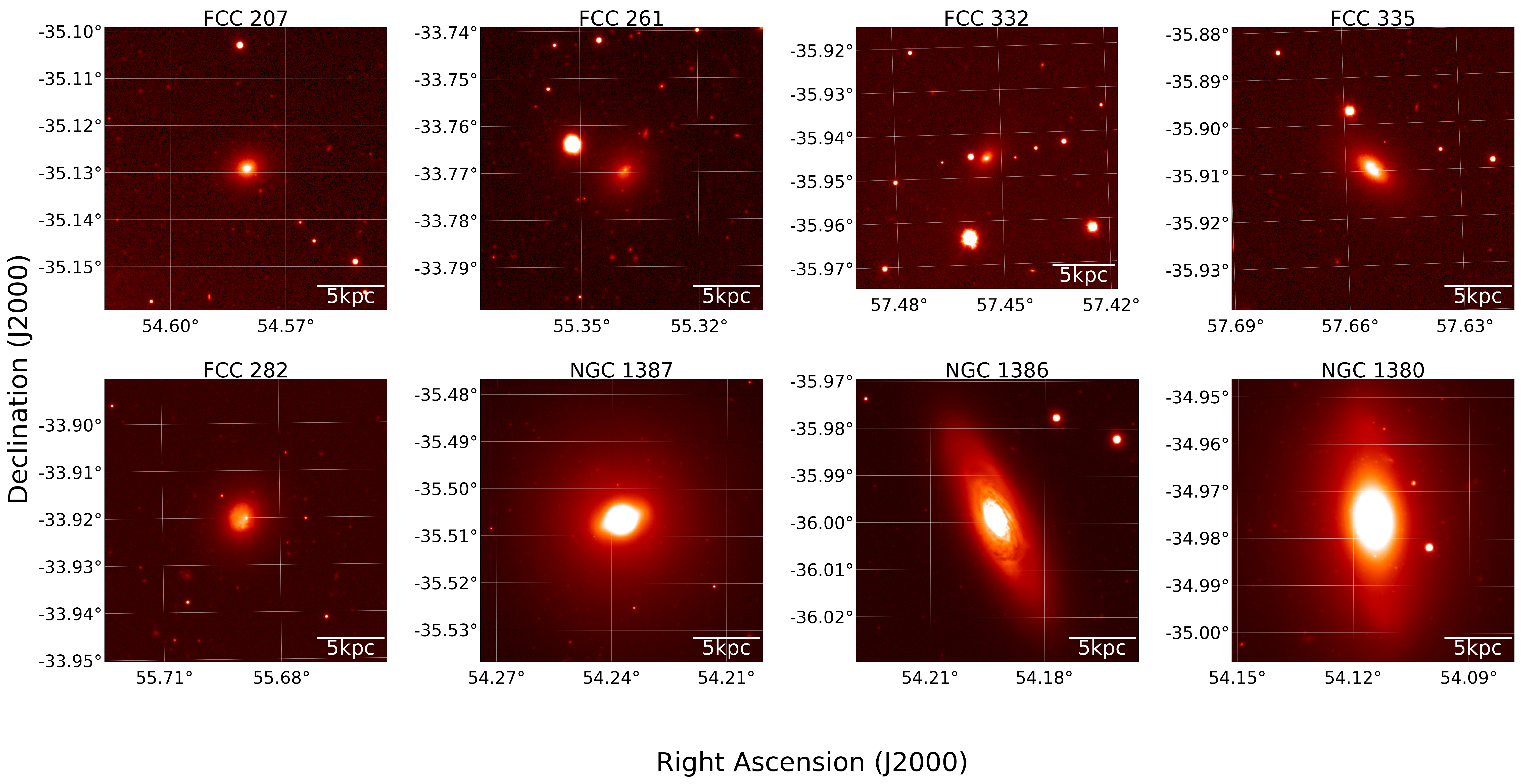}
  \caption{Optical images of the eight \htwo{0pt}-rich \hi{}-undetected galaxies \citep{zabel} for which we evaluated the \mhtwo{}/\mhi{} ratio (Fig.\ref{fig:hionh2}). They are sorted according to increasing \mhi{} upper limit. The \textit{g}-band optical images come from the Fornax Deep Survey (\citealp{FDSIodice2016}; \citealp{venhola2018fds}; \citealp{Reynier_2020arXiv200812633P}). We show a 5~kpc scale bar in the bottom-right corner.}
    \label{fig:nondet_morph}
    \end{figure*}{}   

\end{document}